\documentclass[pdflatex,sn-mathphys-ay]{sn-jnl}% Math and Physical Sciences Author Year Reference Style
\usepackage{graphicx}%
\usepackage{multirow}%
\usepackage{makecell}%
\usepackage{amsmath,amssymb,amsfonts}%
\usepackage{amsthm}%
\usepackage{mathrsfs}%
\usepackage[title]{appendix}%
\usepackage{xcolor}%
\usepackage{textcomp}%
\usepackage{manyfoot}%
\usepackage{booktabs}%
\usepackage{arydshln}%
\usepackage{algorithm}%
\usepackage{algorithmicx}%
\usepackage{algpseudocode}%
\usepackage{listings}%
\usepackage{pgfplots}%
\pgfplotsset{compat=1.18}%
\usetikzlibrary{patterns}%
\usepackage{subcaption}%
\usepackage{float}% For [H] placement option
\usepackage{placeins}% Use \FloatBarrier when needed
\theoremstyle{thmstyleone}%
\theoremstyle{thmstyletwo}%

\theoremstyle{thmstylethree}%

\renewcommand{\orcidlogo}{%
  \includegraphics[width=10pt]{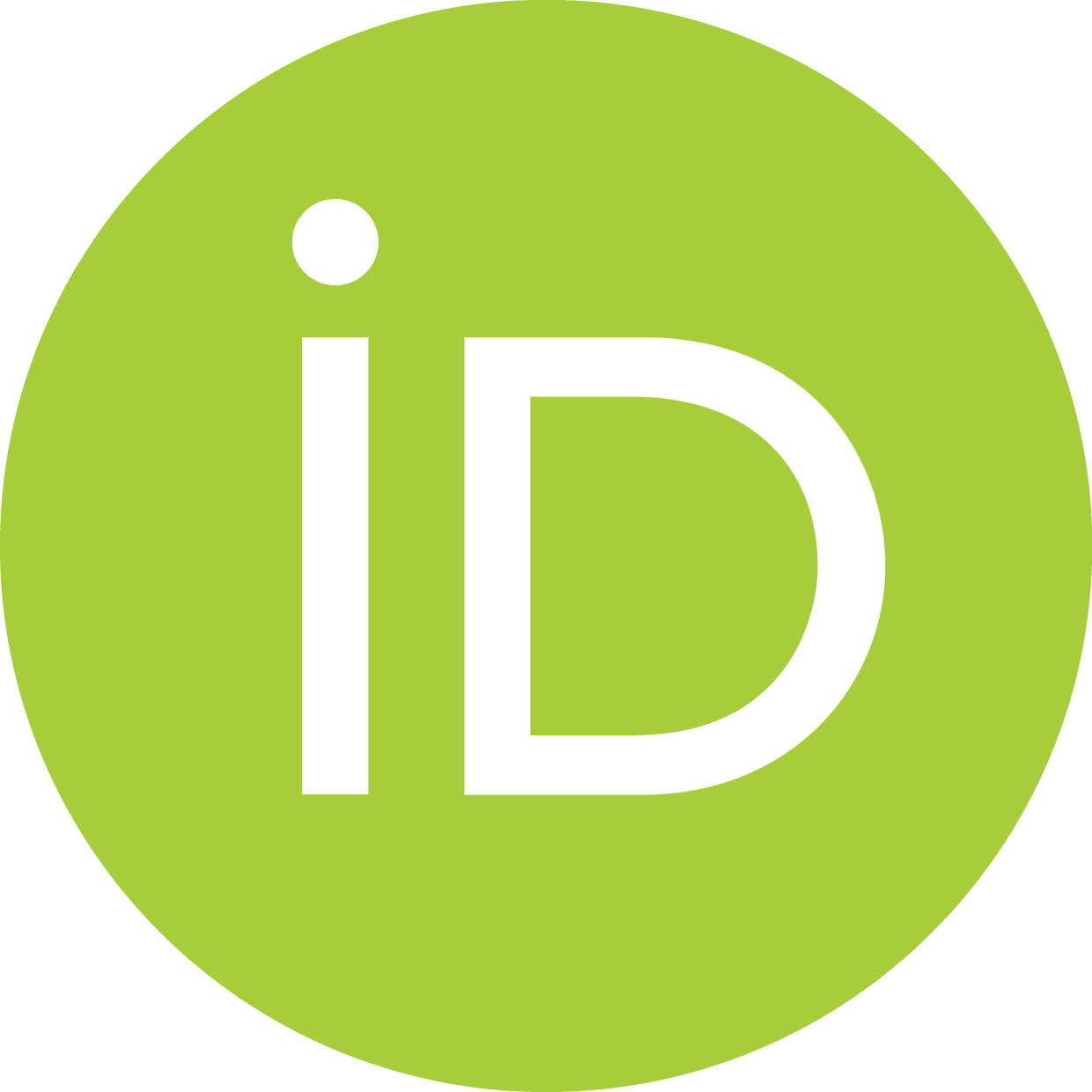}%
}

\begin{document}

\title[Compact SAT Encoding for Power Peak Minimization]{Compact SAT Encoding for Power Peak Minimization in Assembly Line Balancing}

\author[1]{\fnm{Tuyen Van} \sur{Kieu}\email{tuyenkv@vnu.edu.vn}\orcid{https://orcid.org/0009-0007-7800-7172}} 
\author[1]{\fnm{Phong Chi} \sur{Nguyen}\email{24020268@vnu.edu.vn}\orcid{https://orcid.org/0009-0007-0364-2219}}
\author[1]{\fnm{Bao Gia} \sur{Hoang}\email{23020653@vnu.edu.vn}\orcid{https://orcid.org/0009-0004-1400-7358}}
\author*[1]{\fnm{Khanh Van} \sur{To}\email{khanhtv@vnu.edu.vn}\orcid{https://orcid.org/0009-0008-1907-7848}}

\affil[1]{\orgdiv{Faculty of Information Technology}, \orgname{VNU University of Engineering and Technology}, \orgaddress{\city{Hanoi}, \country{Vietnam}}}

\abstract{The Simple Assembly Line Balancing Problem with Power Peak Minimization (SALBP-3PM) minimizes maximum instantaneous power usage while assigning $n$ tasks to $m$ workstations and determining execution schedules within given cycle time constraints. This NP-hard problem couples workstation assignment, temporal sequencing, and power aggregation, presenting significant computational challenges for exact optimization methods. Existing Boolean Satisfiability (SAT) and Maximum Satisfiability (MaxSAT) approaches suffer from baseline encodings generating $O(m^2)$ clauses per precedence edge. We introduce a Compact SAT Encoding (CSE) achieving $O(m)$ clauses per transitive precedence edge using sequential counter techniques. We instantiate four optimization variants: Clause-Based iterative SAT, Pseudo-Boolean (PB) Constraint iterative SAT, MaxSAT, and Incremental SAT. Comprehensive experimental evaluation on benchmark instances demonstrates consistent performance improvements over state-of-the-art approaches, enabling exact optimization on previously intractable industrial-scale instances. The encoding principles generalize to other assembly line balancing variants and broader scheduling problems with precedence constraints.}


\maketitle

\section{Introduction}\label{sec:introduction}
Assembly line balancing (ALB) is a classical topic in production and operations research that seeks to assign tasks to workstations while respecting precedence relations and capacity limits. Canonical formulations such as SALBP-1 and SALBP-2 aim at minimizing the number of stations or the cycle time, respectively, and have motivated a rich body of exact and heuristic methods over several decades \citep{Baybars1986,Scholl2006,scholl_data}. Beyond throughput-centric objectives, modern manufacturing increasingly emphasizes energy-aware design driven by both economic and environmental imperatives. Peak demand charges represent a significant component of industrial electricity bills and are being integrated into production scheduling models \citep{demand_charge_impact,demand_charges}. Many studies also focus on reducing energy costs and emissions through scheduling optimization \citep{manufacturing_energy_cost}. Decision support models that account for energy considerations demonstrate benefits of limiting peak power for grid stability and sustainability \citep{energy_manufacturing}, while manufacturers face mounting pressure to decrease carbon footprints and comply with environmental regulations \citep{Lamy2020IFAC}. In this context, the Simple Assembly Line Balancing Problem with Power Peak Minimization (SALBP-3PM) integrates power consumption into line design by minimizing the maximum instantaneous power usage (the power peak) under given cycle time and number of stations \citep{Gianessi2019SALB3PM}.

The SALBP-3PM was formalized by \citet{Gianessi2019SALB3PM}, who provided an Integer Linear Programming (ILP) model and the first comprehensive computational study. Their approach captures the coupling between assignment, sequencing, and power aggregation over time using time-indexed formulations \citep{pritsker_rcpsp} with big-M constraints \citep{bigm_formulation}. While ILP solvers such as CPLEX \citep{cplex} and Gurobi \citep{gurobi} can solve small to medium instances, scalability becomes challenging as instance size grows, particularly for problems with complex precedence structures and larger time horizons \citep{Gianessi2019SALB3PM,sat_scheduling_survey}.

Satisfiability-based methods have recently emerged as powerful alternatives for SALBP-3PM. \citet{Py2024POS} proposed an iterative SAT framework that progressively refines the search by adding blocking clauses that forbid patterns responsible for high peaks, leveraging the strength of modern conflict-driven clause learning (CDCL) solvers \citep{cdcl,cdcl_modern} such as CaDiCaL \citep{cadical}. In parallel, \citet{Zheng2024MaxSAT} introduced MaxSAT formulations with a compact binary representation of the peak value, demonstrating state-of-the-art performance when coupled with advanced MaxSAT solvers such as EvalMaxSAT \citep{evalmaxsat} and MaxHS \citep{maxhs}. These advances establish SAT/MaxSAT as competitive approaches for peak-oriented line balancing \citep{sat_scalability,sat_scheduling_survey}.

Despite this progress, existing encodings leave room for improvement. Some current encodings contain redundant constraints that inflate formula sizes without contributing to pruning power \citep{encoding_redundancy}. Additionally, the choice between decomposition and global constraints for complex constraints requires trade-offs between compactness and propagation strength \citep{precedence_encoding_comparison}. Moreover, upper bounds on the power peak are typically derived from coarse analytical estimations that can misguide early iterations; bounding techniques in scheduling problems play an important role in accelerating convergence, and similarly for SALBP-3PM, tight upper bound initialization can improve search efficiency \citep{sat_bound_impact}. There remains a need for more compact encodings that better exploit problem structure \citep{compact_encoding_benefits}.

This work addresses these limitations through three integrated contributions. First, we develop a Compact SAT Encoding (CSE) that reduces precedence clause complexity from $O(m^2)$ to $O(m)$ per transitive precedence edge (where $n$ denotes the number of tasks and $m$ the number of workstations) while preserving semantic equivalence. Second, we present a unified optimization framework with four complementary solver variants: clause-based iterative SAT with blocking clauses \citep{Py2024POS}, PB-constraint iterative SAT, MaxSAT combining our compact encoding with the binary peak representation of \citet{Zheng2024MaxSAT}, and incremental SAT with solver state preservation. Third, comprehensive evaluation on 89 benchmark instances demonstrates consistent performance improvements, with our best variant solving all hard instances that timeout competing methods including ILP solvers (CPLEX, Gurobi) and prior SAT/MaxSAT approaches. The encoding principles generalize to other assembly line balancing variants and broader scheduling problems with precedence constraints \citep{Baybars1986,Scholl2006,pritsker_rcpsp,sat_scheduling_survey}.

The remainder of this paper is organized as follows: Section~\ref{sec:preliminaries} formalizes SALBP-3PM. Section~\ref{sec:related_work} reviews related work and existing SAT/MaxSAT encodings. Section~\ref{sec:proposed_model} details our Compact SAT Encoding with theoretical efficiency analysis and four optimization variants. Section~\ref{sec:experiments} describes the experimental methodology and benchmark instances. Section~\ref{sec:results} reports comprehensive results comparing our approaches against state-of-the-art methods. Section~\ref{sec:conclusion} concludes with summary, limitations, and future research directions. Appendices provide detailed theoretical analysis, experimental tables, and algorithmic descriptions.

\section{Preliminaries}\label{sec:preliminaries}

\subsection{Problem Definition}
We formalize the Simple Assembly Line Balancing Problem with Power Peak Minimization (SALBP-3PM). Let $N=\{1,\ldots,n\}$ be the finite set of tasks. Each task $i\in N$ has (i) an integer processing time $t_i\in \mathbb{N}_{>0}$ (measured in discrete time units) and (ii) a constant power consumption rate $w_i\in \mathbb{N}_{>0}$ while it is being processed. Let $S=\{1,\ldots,m\}$ denote the (fixed) set of workstations and let $c\in \mathbb{N}_{>0}$ be the cycle time. Time is discretized into the horizon $T=\{0,\ldots,c-1\}$. For each task $i$, the admissible start times are $T^i=\{0,\ldots,c-t_i\}$. Precedence relations between tasks are given by a directed acyclic graph (DAG) $G=(N,P)$ where each arc $(i,j)\in P$ enforces that $i$ must be completed before $j$ starts \citep{Gianessi2019SALB3PM}.

We adopt a scheduling (time-indexed) view in which a \emph{solution} consists of: (i) an assignment of each task $i$ to exactly one workstation $a(i)\in S$, and (ii) an integer start time $\sigma_i\in T^i$. Task $i$ then occupies (is processed during) the interval $[\sigma_i,\sigma_i+t_i-1]\subseteq T$ and contributes its power $w_i$ to every time unit in that interval.

Idle time \emph{within} a workstation is allowed in the general SALBP-3PM (distinguishing it from the no-idle variant studied by \citet{Lamy2020IFAC}); sequencing is therefore an integral part of the decision since it influences the temporal superposition of power consumptions across stations.

Let $A(\tau)=\{ i\in N \mid \sigma_i \le \tau < \sigma_i + t_i \}$ be the set of tasks active at (discrete) time $\tau\in T$. The instantaneous total power at time $\tau$ induced by a solution $(a,\sigma)$ is
\begin{equation}
W(\tau)= \sum_{i \in A(\tau)} w_i.\label{eq:instant_power}
\end{equation}
The objective of SALBP-3PM is to minimize the peak (maximum) instantaneous power over the cycle:
\begin{equation}
W^{*} = \min_{(a,\sigma)\ \text{feasible}}\ \max_{\tau \in T} W(\tau).\label{eq:objective}
\end{equation}

Feasibility requires the following constraints:
\begin{enumerate}
    \item \emph{Task assignment:} Each task must be processed by exactly one workstation, i.e., $a: N \to S$ is a total function.
    \item \emph{Scheduling:} Each task must start within its admissible time window, i.e., $\sigma_i \in T^i$ for all $i \in N$.
    \item \emph{Non-overlap:} Tasks assigned to the same workstation cannot overlap temporally. Formally, if $a(i)=a(j)$ and $i\neq j$, then either $\sigma_i + t_i \le \sigma_j$ or $\sigma_j + t_j \le \sigma_i$.
    \item \emph{Precedence:} For every precedence relation $(i,j)\in P$, task $i$ must complete before task $j$ starts. This is satisfied if either (a) $a(i)<a(j)$, placing task $i$ at an earlier-indexed workstation, or (b) $a(i)=a(j)$ and $\sigma_i + t_i \le \sigma_j$, ensuring temporal ordering within the same workstation.
    \item \emph{Cycle time:} All task processing must complete within the cycle horizon, i.e., $\sigma_i + t_i \le c$ for all $i \in N$. Note that this constraint is implicitly satisfied when $\sigma_i\in T^i$, since $T^i=\{0,\ldots,c-t_i\}$.
\end{enumerate}

Because adding start-time decisions generalizes classical SALBP variants (e.g., SALBP-1 and SALBP-2 \citep{Baybars1986,Scholl2006}), SALBP-3PM is NP-hard. The power peak component further couples balancing and sequencing, producing a combinatorial search space that challenged early exact (ILP) formulations \citep{Gianessi2019SALB3PM}.

Table~\ref{tab:notation} summarizes the notation used throughout this paper for clarity and consistency.

\begin{table}[!htb]
\centering
\caption{Notation and variable definitions for SALBP-3PM}
\label{tab:notation}
\small
\begin{tabular}{@{}ll@{}}
\toprule
\textbf{Notation} & \textbf{Description} \\
\midrule
\multicolumn{2}{l}{\textit{Problem Parameters}} \\
$n$ & Number of tasks \\
$m$ & Number of workstations \\
$c$ & Cycle time (time horizon length) \\
$N = \{1,\ldots,n\}$ & Set of tasks \\
$S = \{1,\ldots,m\}$ & Set of workstations \\
$T = \{0,\ldots,c-1\}$ & Discrete time horizon \\
$t_i$ & Processing time of task $i$ \\
$\bar{t}$ & Average processing time: $\bar{t} = \frac{1}{n}\sum_{i=1}^{n} t_i$ \\
$w_i$ & Power consumption rate of task $i$ \\
$T^i = \{0,\ldots,c-t_i\}$ & Admissible start times for task $i$ \\
$P \subseteq N \times N$ & Precedence constraints (directed edges) \\
$E$ & Original precedence edge set (same as $P$) \\
$E^*$ & Extended precedence graph (transitive closure of $E$) \\
$i \prec j$ & Task $i$ must complete before task $j$ starts \\
\midrule
\multicolumn{2}{l}{\textit{Solution Components}} \\
$a(i) \in S$ & Workstation assignment for task $i$ \\
$\sigma_i \in T^i$ & Start time of task $i$ \\
$A(\tau)$ & Set of tasks active at time $\tau$ \\
$W(\tau)$ & Total instantaneous power at time $\tau$: $\sum_{i \in A(\tau)} w_i$ \\
$W^*$ & Objective: minimum power peak (to minimize) \\
\midrule
\multicolumn{2}{l}{\textit{SAT/MaxSAT Variables}} \\
$X_{i,k}$ & Boolean: task $i$ assigned to workstation $k$ \\
$S_{i,t}$ & Boolean: task $i$ starts at time $t$ \\
$A_{i,t}$ & Boolean: task $i$ is active (running) at time $t$ \\
$U_j$ & Peak indicator: $U_j = 1$ iff peak $\geq j$ \citep{Zheng2024MaxSAT} \\
$\text{binU}_b$ & Binary peak bit: $W_{peak} = \sum_b 2^b \cdot \text{binU}_b$ \citep{Zheng2024MaxSAT} \\
\midrule
\multicolumn{2}{l}{\textit{Auxiliary Variables (Compact SAT Encoding)}} \\
$R_{j,k}$ & Boolean: task $j$ assigned to workstation $\leq k$ \\
$T_{j,\tau}$ & Boolean: task $j$ starts at time $\leq \tau$ \\
\bottomrule
\end{tabular}
\end{table}

\begin{table}[!htb]
\centering
\caption{Processing time and power consumption of each task in the illustrative example.}
\label{tab:example_data}
\begin{tabular}{@{}lccccc@{}}
\toprule
Task & 1 & 2 & 3 & 4 & 5 \\
\midrule
Processing time $t_i$ & 3 & 4 & 2 & 3 & 2 \\
Power consumption $w_i$ & 5 & 3 & 6 & 4 & 5 \\
\bottomrule
\end{tabular}
\end{table}

\begin{figure}[!htb]
\centering
\includegraphics[width=\textwidth]{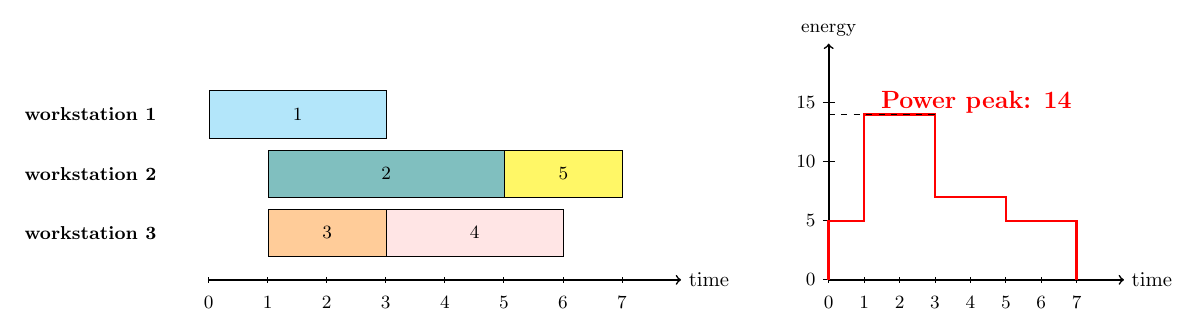}
\caption{Illustrative example of SALBP-3PM with 5 tasks distributed across 3 workstations over a cycle time of 7 units}
\label{fig:salbp3pm_example}
\end{figure}

To illustrate the problem, consider the example shown in Figure~\ref{fig:salbp3pm_example} with task data from Table~\ref{tab:example_data}. The left panel displays the schedule visualization across three workstations: task 1 executes in interval $[0,3)$ on workstation 1; tasks 2 and 5 occupy intervals $[1,5)$ and $[5,7)$ respectively on workstation 2; tasks 3 and 4 are assigned to workstation 3 with intervals $[1,3)$ and $[3,6)$. The right panel shows the resulting power profile over time. At time $\tau=1$, three tasks (1, 2, and 3) execute simultaneously, producing the peak power $W^{*}=w_1+w_2+w_3=5+3+6=14$. The objective is to minimize this peak through optimal task assignment and temporal sequencing decisions.

Two immediate lower bounds on the optimum peak $W^{*}$ are: (i) $\max_{i \in N} w_i$ (any feasible solution must at least accommodate the largest single-task power), and (ii) the average power $\left\lceil \dfrac{\sum_{i \in N} w_i t_i}{c} \right\rceil$, since the total ``energy'' $\sum_i w_i t_i$ is spread over $c$ time units. A trivial upper bound is $\sum_{i \in N} w_i$. Tighter bounds can be obtained through constructive heuristics or (as proposed in this work) an initial SAT feasibility solve that yields a concrete peak value. Prior approaches often relied on coarse analytical upper bounds, which may slow optimization \citep{Py2024POS,Zheng2024MaxSAT}.

Following the established benchmark protocol \citep{Gianessi2019SALB3PM,Py2024POS,Zheng2024MaxSAT}, precedence graphs and cycle times are adopted from standard SALBP data sets \citep{scholl_data,otto_survey}. The number of stations $m$ is fixed to the optimal SALBP-1 value $m^{*}$, ensuring feasibility under the given cycle time (i.e., a solution exists achieving some peak). This setting isolates the impact of sequencing and temporal aggregation on $W^{*}$.

Given $(N,S,G,c,(t_i)_{i\in N},(w_i)_{i\in N})$, the problem seeks $a$ and $\sigma$ satisfying constraints (1)--(5) that minimize the objective in Eq.~\eqref{eq:objective}.

This formalization serves as the basis for the compact SAT encoding and optimization schemes developed in the remainder of the paper, extending and refining prior ILP, SAT, and MaxSAT models \citep{Gianessi2019SALB3PM,Lamy2020IFAC,Py2024POS,Zheng2024MaxSAT}.

\section{Related Work and Background}\label{sec:related_work}

\subsection{Related Work}
\label{subsec:related_work}
The integration of power peak minimization into simple assembly line balancing extends the classical SALBP family \citep{Baybars1986,Scholl2006,scholl_data} by coupling workstation assignment with intra-station temporal sequencing. \citet{Gianessi2019SALB3PM} established SALBP-3PM and provided the first integer linear programming (ILP) formulation using time-indexed variables and big-$M$ linearization \citep{bigm_formulation}. While their CPLEX-based approach \citep{cplex} solved small and medium instances, performance degraded with dense precedence graphs and high cycle slack, reflecting combinatorial challenges in time-indexed scheduling formulations \citep{pritsker_rcpsp,sat_scheduling_survey}.

\citet{Lamy2020IFAC} examined a no-idle variant where tasks process contiguously within stations, developing a metaheuristic (MSxELS) that demonstrated practical relevance for reducing energy costs. Studies on production scheduling with time-of-use electricity pricing and peak demand charges \citep{demand_charge_impact} and energy-efficient manufacturing scheduling \citep{manufacturing_energy_cost} further highlight the importance of energy-aware optimization. However, metaheuristic approaches lack optimality guarantees.

Satisfiability-based methods recently emerged as powerful alternatives. \citet{Py2024POS} proposed an iterative SAT framework decoupling feasibility from optimization. Their approach solves base constraints to find feasible schedules, then iteratively adds blocking clauses to exclude peak-inducing task configurations until UNSAT certifies optimality. Leveraging CDCL solvers \citep{cdcl_modern,cadical}, they outperformed ILP on several instances, though clause accumulation and encoding inefficiencies limited scalability. Redundant constraints in clause learning \citep{encoding_redundancy} and decomposition choices for complex constraints \citep{precedence_encoding_comparison} affect formula sizes and solver performance.

\citet{Zheng2024MaxSAT} introduced MaxSAT formulations with binary (logarithmic) peak encoding \citep{compact_encoding_benefits}, reducing variables from $O(UB)$ to $O(\log UB)$ where $UB$ is the peak upper bound. State-of-the-art solvers (EvalMaxSAT \citep{evalmaxsat}, MaxHS \citep{maxhs}, RC2 \citep{rc2}) exploited this compact representation for improved robustness \citep{maxsat_robustness}. However, bounding techniques in scheduling play a crucial role; as shown for resource-constrained project scheduling \citep{sat_bound_impact}, tight bounds accelerate convergence, yet analytical upper bounds for SALBP-3PM remained conservative.

These works reveal methodological gaps: (i) existing encodings use $O(m^2)$ clauses per precedence edge rather than $O(m)$ compact alternatives \citep{compact_encoding_benefits}, and redundant constraints and decomposition strategies affect encoding size \citep{encoding_redundancy,precedence_encoding_comparison}; (ii) bounding techniques are important in scheduling \citep{sat_bound_impact}, yet loose analytical bounds delay pruning; (iii) clause-based versus PB-based iterative refinement lack systematic comparison \citep{Py2024POS,Zheng2024MaxSAT}; and (iv) incremental solver reuse across iterations remains underexplored \citep{cdcl_modern,sat_scalability}. Building on the foundations established by previous work \citep{Py2024POS,Zheng2024MaxSAT}, we introduce five key contributions: (i) a compact encoding achieving $O(m)$ precedence clauses using sequential counter techniques \citep{sinz_sequential}; (ii) tight bounds obtained from a preliminary initialization phase; (iii) four unified optimization variants (clause-based, PB-constraint, MaxSAT, and incremental SAT); (iv) extended precedence preprocessing; and (v) constraint refinements that remove redundancy. This integrated contribution advances both encoding efficiency and methodological breadth for SALBP-3PM.

\subsection{Existing SAT and MaxSAT Encodings}\label{subsec:existing_encodings}

The integration of satisfiability-based methods into SALBP-3PM has yielded two complementary paradigms: iterative SAT approaches that decompose optimization into sequential feasibility problems, and direct MaxSAT formulations that express the peak objective through weighted soft clauses. This subsection presents the foundational encodings from prior work \citep{Py2024POS,Zheng2024MaxSAT}, establishing the baseline upon which our compact reformulation builds.

\subsubsection{SAT-based Iterative Framework}
\label{subsubsec:sat_baseline}

The SAT-based approach \citep{Py2024POS} models SALBP-3PM through Boolean decision variables and iterative refinement. The encoding employs three variable types: $X_{i,k}$ (task $i$ assigned to workstation $k$), $S_{i,t}$ (task $i$ starts at time $t$), and $A_{i,t}$ (task $i$ active at time $t$).

The feasibility encoding comprises structural constraints governing task assignment, precedence relationships, scheduling requirements, activity propagation, and non-overlap enforcement.

\emph{Task Assignment:}
\begin{align}
\bigvee_{k \in S} X_{i,k} &\quad \forall i \in N \tag{SAT-1} \\
\overline{X_{i,k_1}} \vee \overline{X_{i,k_2}} &\quad \forall i \in N, k_1, k_2 \in S: k_1 < k_2 \tag{SAT-2}
\end{align}

\emph{Precedence:}
\begin{align}
\overline{X_{i,k}} \vee \overline{X_{j,h}} &\quad \forall i,j \in N: i \prec j, k,h \in S: k > h \tag{SAT-3} \\
\overline{X_{i,k}} \vee \overline{X_{j,k}} \vee \overline{S_{i,t_1}} \vee \overline{S_{j,t_2}} &\quad \forall i,j \in N: i \prec j, k \in S, t_1 \in T^i, t_2 \in T^j: t_1 > t_2 \tag{SAT-4}
\end{align}

\emph{Scheduling:}
\begin{align}
\bigvee_{t \in T^i} S_{i,t} &\quad \forall i \in N \tag{SAT-5} \\
\overline{S_{i,t_1}} \vee \overline{S_{i,t_2}} &\quad \forall i \in N, t_1, t_2 \in T^i: t_1 < t_2 \tag{SAT-6} \\
\overline{S_{i,t}} &\quad \forall i \in N, t \in T: t \notin T^i \tag{SAT-7}
\end{align}

\emph{Activity and Non-overlap:}
\begin{align}
\overline{S_{i,t}} \vee A_{i,t+\epsilon} &\quad \forall i \in N, t \in T^i, \epsilon \in [0, t_i-1] \tag{SAT-8} \\
\overline{X_{i,k}} \vee \overline{X_{j,k}} \vee \overline{A_{i,t}} \vee \overline{A_{j,t}} &\quad \forall i,j \in N: i \neq j, k \in S, t \in T \tag{SAT-9}
\end{align}

\emph{Simplification (preprocessing-based pruning):}
\begin{align}
\overline{X_{i,k}} &\quad \forall i \in N, k \in S: \text{ip}(i,k) \tag{SAT-10} \\
\overline{X_{i,k}} \vee \overline{S_{i,t}} &\quad \forall i \in N, k \in S, t \in T^i: \text{ip}(i,k,t) \tag{SAT-11} \\
\overline{A_{i,t}} &\quad \forall i \in N, t \in [c-t_i, t_i-1] \cap T \tag{SAT-12}
\end{align}
where $\text{ip}(i,k)$ and $\text{ip}(i,k,t)$ denote infeasible assignments identified through precedence and temporal analysis.

Peak minimization is achieved iteratively by adding blocking clauses that eliminate previously discovered peak configurations:
\begin{align}
\bigvee_{i \in C} \overline{A_{i,t}} &\quad \forall C \in \mathbb{C}, t \in T \tag{SAT-13}
\end{align}
where $\mathbb{C}$ collects task sets causing peak configurations. Initially empty, $\mathbb{C}$ expands after each SAT solve until UNSAT certifies optimality.

\subsubsection{MaxSAT-based Direct Optimization}
\label{subsubsec:maxsat_baseline}

The MaxSAT formulations \citep{Zheng2024MaxSAT} encode peak minimization directly through weighted soft clauses. Two variants are proposed: an inaugural model using peak indicator variables $U_j$ and an enhanced model employing binary peak representation with $\text{binU}_b$.

The inaugural MaxSAT model extends the SAT baseline with peak indicators $U_j$ for $j \in \{1, \ldots, UB\}$ where $UB = \sum_{w_i \in W_m} w_i$ (sum of $m$ highest powers), defining the peak value as $W_{peak} = \sum_{j=1}^{UB} U_j$.

\emph{Soft constraints:}
\begin{equation}
\lnot U_j \quad \forall j \in \{1, \ldots, UB\} \tag{MaxSAT-1}
\end{equation}

\emph{Hard constraints:}
The feasibility constraints (MaxSAT-2 through MaxSAT-8) mirror the SAT baseline (SAT-1 through SAT-9), enforcing task assignment uniqueness, precedence relationships, scheduling requirements, activity propagation, and non-overlap. The key additions are peak-related constraints:
\begin{align}
U_j &\quad \forall j \in \{1, \ldots, LB\} \tag{MaxSAT-9}\\
\lnot U_j \lor U_{j-1} &\quad \forall j \in \{2, \ldots, UB\} \tag{MaxSAT-10}\\
\sum_{i=1}^n w_i \cdot A_{i,t} + \sum_{j=1}^{UB} \lnot U_j &\leq UB \quad \forall t \in T \tag{MaxSAT-11}
\end{align}
where $LB = \max_{i \in N} w_i$ establishes the lower bound on feasible peaks.

To reduce peak indicator variables from $UB$ (potentially large) to $\lceil \log_2 UB \rceil$, the binary representation enhancement replaces $U_j$ with binary bit indicators $binU_j$ for $j \in \{0, \ldots, \lceil \log_2 UB \rceil\}$, yielding $W_{peak} = \sum_{j=0}^{\lceil \log_2 UB \rceil} 2^j \cdot binU_j$. This approach draws on cardinality constraint encodings (sequential counter \citep{sinz_sequential}, totalizer \citep{totalizer}, cardinality networks \citep{cardinality_networks}), adapted for peak representation:

\emph{Soft constraints:}
\begin{equation}
(\lnot binU_j, 2^j) \quad \forall j \in \{0, \ldots, \lceil \log_2 UB \rceil\} \tag{MaxSAT-1'}
\end{equation}

\emph{Modified hard constraints (feasibility constraints unchanged):}
\begin{align}
\sum_{j=0}^{\lceil \log_2 UB \rceil} 2^j \cdot binU_j &\geq LB \tag{MaxSAT-9'}\\
\sum_{i=1}^n w_i \cdot A_{i,t} + \sum_{j=0}^{\lceil \log_2 UB \rceil} 2^j \cdot \lnot binU_j &\leq UB' \quad \forall t \in T \tag{MaxSAT-11'}
\end{align}
where $UB' = \sum_{j=0}^{\lceil \log_2 UB \rceil} 2^j$.

Zheng et al.'s empirical evaluation demonstrated that the binary model with EvalMaxSAT solver achieved superior performance compared to the inaugural model, particularly on larger instances where the logarithmic variable reduction becomes significant \citep{Zheng2024MaxSAT}. Following these findings, our work adopts the binary peak encoding for MaxSAT formulations, combining it with our Compact SAT Encoding (Section~\ref{subsec:compact_encoding}).

\subsection{Motivation for Compact Encoding}
\label{subsec:limitations}

While these baseline encodings have demonstrated effectiveness, several structural inefficiencies motivate refinement. First, precedence constraints (SAT-3, SAT-4) generate $O(m^2)$ clauses per edge, inflating formula size. Second, the upper bound $UB = \sum_{w_i \in W_m} w_i$ significantly exceeds true optima for sparse precedence graphs, producing many vacuous soft clauses in MaxSAT formulations. Third, iterative SAT rebuilds solver state from scratch at each iteration, foregoing opportunities for incremental reasoning about learned conflict clauses. These observations guide the compact encoding and optimization enhancements presented in Section~\ref{sec:proposed_model}.

\section{Compact SAT Encoding and Optimization Approaches}\label{sec:proposed_model}

This section presents our compact encoding approach that addresses the scalability limitations of baseline formulations. We build upon the sequential counter encoding technique \citep{sinz_sequential} that uses cumulative reachability variables to achieve $O(m)$ clause complexity per precedence constraint (versus $O(m^2)$ for baseline SAT encodings \citep{Py2024POS,Zheng2024MaxSAT}), and extended precedence graph preprocessing \citep{bofill2017smt} that identifies implied constraints through transitive closure analysis.

Our contributions extend these techniques in three ways. First, we integrate the sequential counter encoding with forward-backward temporal analysis \citep{Py2024POS} to systematically eliminate infeasible variable assignments, significantly reducing active constraint ranges for typical instances. Second, we develop an enhanced upper bound initialization strategy using a multi-solve phase with solution blocking, providing tighter bounds than analytical estimates and reducing optimization iterations. Third, we propose four algorithmic variants—Clause-Based, PB-Constraint, MaxSAT (leveraging the binary peak representation of \citet{Zheng2024MaxSAT}), and Incremental SAT—each suited to different problem characteristics while sharing the compact encoding foundation.

\subsection{Compact Precedence Encoding}
\label{subsec:compact_encoding}

The compact encoding derives from a sequence-based formulation of precedence constraints. Given precedence constraint $i \prec j$ and workstation set $S = \{1, 2, \ldots, m\}$, the precedence requirement can be expressed as:
\begin{equation}
\sum_{k=1}^{\ell} X_{j,k} + X_{i,\ell+1} \leq 1 \quad \forall \ell \in \{1, 2, \ldots, m-1\}
\label{eq:sequence_form}
\end{equation}
This formulation captures the essential precedence requirement: if task $i$ is assigned to station $\ell+1$, then task $j$ cannot be assigned to any station in $\{1, \ldots, \ell\}$. The sequence of constraints~\eqref{eq:sequence_form} exhibits a staircase structure where consecutive constraints share prefix sums $\sum_{k=1}^{\ell} X_{j,k}$. By introducing cumulative reachability variables, we decompose each constraint into binary clauses via sequential counter encoding \citep{sinz_sequential}.

\subsubsection{Variable Definitions}

Beyond the base variables ($X_{i,k}$: task-workstation assignment, $S_{i,t}$: task start time, $A_{i,t}$: task active at time $t$), we introduce cumulative reachability variables:
\begin{itemize}
\item $R_{i,k}$: true if task $i$ assigned to workstation $k$ or earlier, i.e., $R_{i,k} \equiv \bigvee_{\ell=1}^{k} X_{i,\ell}$
\item $T_{i,t}$: true if task $i$ started by time $t$ or earlier, i.e., $T_{i,t} \equiv \bigvee_{\tau=0}^{t} S_{i,\tau}$
\end{itemize}
These variables satisfy monotonicity ($R_{i,k} \Rightarrow R_{i,k+1}$ and $T_{i,t} \Rightarrow T_{i,t+1}$), enabling compact precedence encoding through sequential counter techniques \citep{sinz_sequential}.

\subsubsection{Constraint Formulation}

Our encoding combines compact constraints (CSE-1 through CSE-11) for workstation assignment and temporal scheduling with direct constraints (CSE-12 through CSE-16) for activity tracking and preprocessing-based simplifications. Several constraints are directly reused from prior work \citep{Py2024POS,Zheng2024MaxSAT}: CSE-12 (activity propagation, equivalent to SAT-8), CSE-14 (non-overlap, equivalent to SAT-9), CSE-15 (workstation pruning, equivalent to SAT-10), and CSE-16 (temporal pruning, equivalent to SAT-11). The key innovations are the compact at-most-one (AMO) and precedence constraints (CSE-1--6, CSE-7--11, CSE-13) that achieve quadratic-to-linear complexity reduction through sequential counter encoding. The complete formulation is organized as follows.

The workstation assignment uses cumulative reachability through sequential constraints:
\begin{align}
R_{i,1} \Leftrightarrow X_{i,1} &\quad \forall i \in N \tag{CSE-1}\\
R_{i,k-1} \Rightarrow R_{i,k} &\quad \forall i \in N, k \in \{2, \ldots, m\} \tag{CSE-2} \\
X_{i,k} \Rightarrow R_{i,k} &\quad \forall i \in N, k \in \{1, \ldots, m\} \tag{CSE-3} \\
X_{i,k} \Rightarrow \overline{R_{i,k-1}} &\quad \forall i \in N, k \in \{2, \ldots, m\} \tag{CSE-4} \\
(R_{i,k} \land \overline{R_{i,k-1}}) \Rightarrow X_{i,k} &\quad \forall i \in N, k \in \{2, \ldots, m\} \tag{CSE-5}\\
R_{i,m-1} \vee X_{i,m} &\quad \forall i \in N \tag{CSE-5a}\\
X_{i,k+1} \Rightarrow R_{j,k} &\quad \forall (i,j) \in E^*, k \in \{1, \ldots, m-1\} \tag{CSE-6}
\end{align}
Constraint CSE-6 enforces precedence: when task $i$ is assigned to workstation $k+1$, its predecessor $j$ (where $i \prec j$ in precedence) must have been assigned to a workstation $\leq k$. Equivalently, task $i$ cannot be at workstation $k+1$ unless predecessor $j$ has reached workstation $k$ or earlier. This achieves $O(m)$ clauses per precedence edge versus $O(m^2)$ for baseline pairwise encodings (SAT-3).

Temporal scheduling employs an analogous structure for start time variables:
\begin{align}
T_{i,0} \Leftrightarrow S_{i,0} &\quad \forall i \in N \tag{CSE-7}\\
T_{i,t-1} \Rightarrow T_{i,t} &\quad \forall i \in N, t \in \{1, \ldots, c-t_i\} \tag{CSE-8} \\
S_{i,t} \Rightarrow T_{i,t} &\quad \forall i \in N, t \in \{0, \ldots, c-t_i\} \tag{CSE-9} \\
S_{i,t} \Rightarrow \overline{T_{i,t-1}} &\quad \forall i \in N, t \in \{1, \ldots, c-t_i\} \tag{CSE-10} \\
(T_{i,t} \land \overline{T_{i,t-1}}) \Rightarrow S_{i,t} &\quad \forall i \in N, t \in \{1, \ldots, c-t_i\} \tag{CSE-11}
\end{align}

Activity tracking and resource allocation constraints complete the encoding:
\begin{align}
S_{i,t} \Rightarrow A_{i,t+\ell} &\quad \forall i \in N, t \in \{0, \ldots, c-t_i\}, \ell \in \{0, \ldots, t_i-1\} \tag{CSE-12}\\
(X_{i,k} \land X_{j,k} \land T_{j,t}) \Rightarrow \overline{S_{i,t'}} &\quad \forall (i,j) \in E^*, k \in S, t, t' \in T: t' \leq t + t_j \tag{CSE-13}\\
(X_{i,k} \land X_{j,k}) \Rightarrow \overline{(A_{i,t} \land A_{j,t})} &\quad \forall i,j \in N: i \neq j, k \in S, t \in T \tag{CSE-14}\\
\overline{X_{i,k}} &\quad \forall i \in N, k \in S: \text{First}(i) > k \vee \text{Last}(i) < k \tag{CSE-15} \\
X_{i,k} \Rightarrow \overline{S_{i,t}} &\quad \forall i \in N, k \in S, t \in T: \text{ip}(i,k,t) = \text{true} \tag{CSE-16}
\end{align}
Constraint CSE-12 ensures that a task remains active throughout its execution duration. Constraint CSE-13 enforces temporal precedence within the same workstation, using the cumulative variable $T_{j,t}$ for compactness. Constraint CSE-14 prevents overlapping execution of distinct tasks within the same workstation. Constraints CSE-15 and CSE-16 eliminate infeasible assignments identified during preprocessing, where $\text{First}(i)$ and $\text{Last}(i)$ denote the earliest and latest feasible workstations for task $i$, and $\text{ip}(i,k,t)$ indicates temporal infeasibility.

\subsubsection{Complexity Analysis}

Tables~\ref{tab:variable_comparison} and~\ref{tab:complexity_comparison} summarize the variable and clause complexity comparisons between ORG encoding (Section~\ref{subsubsec:sat_baseline}) and CSE encoding.

\begin{table}[!htb]
\caption{Variable complexity comparison between ORG and CSE encodings.}
\label{tab:variable_comparison}
\small
\centering
\begin{tabular}{@{}lll@{}}
\toprule
\textbf{Variable Type} & \textbf{ORG} & \textbf{CSE} \\
\midrule
Task assignment $X_{i,k}$ & $O(nm)$ & $O(nm)$ \\
Start time $S_{i,t}$ & $O(nc)$ & $O(nc)$ \\
Activity $A_{i,t}$ & $O(nc)$ & $O(nc)$ \\
Reach (station) $R_{i,k}^\dagger$ & --- & $O(nm)$ \\
Reach (time) $T_{i,t}^\dagger$ & --- & $O(nc)$ \\
Peak indicator $U_j^\ddagger$ & $O(UB)$ & $O(UB_{tight} - LB)$ \\
Binary peak $binU_b^\ddagger$ & $O(\log_2 UB)$ & $O(\log_2 UB_{tight})$ \\
\midrule
\textbf{Total} & $O(nm + nc + \log UB)$ & $O(2nm + 2nc + \log UB_{tight})$ \\
\bottomrule
\multicolumn{3}{l}{\footnotesize $^\dagger$Auxiliary variables. $^\ddagger$Peak variables: $U_j$ for CSE\_INC, $\text{binU}_b$ for CSE\_MaxHS/CSE\_Eval.}
\end{tabular}
\end{table}

\begin{table}[!htb]
\caption{Clause complexity comparison between ORG and CSE encodings. Compact constraints are marked with $\star$.}
\label{tab:complexity_comparison}
\small
\centering
\begin{tabular}{@{}lll@{}}
\toprule
\textbf{Constraint Type} & \textbf{ORG} & \textbf{CSE} \\
\midrule
Task assignment ALO & SAT-1: $O(n)$ & CSE-1,5a: $O(n)$ \\
Task assignment AMO$^\star$ & SAT-2: $O(nm^2)$ & CSE-2,3,4,5: $\mathbf{O(nm)}$ \\
Scheduling ALO & SAT-5: $O(n)$ & CSE-7: $O(n)$ \\
Scheduling AMO$^\star$ & SAT-6: $O(nc^2)$ & CSE-8,9,10,11: $\mathbf{O(nc)}$ \\
Precedence (diff. station)$^\star$ & SAT-3: $O(|E|m^2)$ & CSE-6: $\mathbf{O(|E^*|m)}$ \\
Precedence (same station)$^\star$ & SAT-4: $O(|E|mc^2)$ & CSE-13: $\mathbf{O(|E^*|mc)}$ \\
Activity propagation & SAT-8: $O(nc\bar{t})$ & CSE-12: $O(nc\bar{t})$ \\
Non-overlap & SAT-9: $O(n^2mc)$ & CSE-14: $O(n^2mc)$ \\
\midrule
\textbf{Total (dominant terms)} & $O(nm^2 + nc^2 + |E|mc^2)$ & $\mathbf{O(nm + nc + |E^*|mc)}$ \\
\bottomrule
\end{tabular}
\end{table}

While CSE introduces $O(nm + nc)$ auxiliary variables ($R_{i,k}$ and $T_{i,t}$), the use of $UB_{tight}$ instead of analytical $UB$ compensates by reducing binary peak variables from $O(\log_2 UB)$ to $O(\log_2 UB_{tight})$. Since $UB_{tight} < UB$ (as demonstrated empirically in Section~\ref{subsec:encoding_size}), this partially offsets the auxiliary variable overhead while simultaneously achieving $O(nm^2 + nc^2) \to O(nm + nc)$ clause reduction. The auxiliary reach variables also enhance unit propagation efficiency through binary clause chains with clear semantic meaning (cumulative reachability).

The key insight is that sequential counter encoding \citep{sinz_sequential} replaces pairwise at-most-one (AMO) constraints with linear-size formulations using cumulative variables ($R_{i,k}$ for workstations, $T_{i,t}$ for time). For task assignment, ORG's SAT-2 requires $n \cdot \binom{m}{2} = O(nm^2)$ pairwise clauses, while CSE's cumulative constraints (CSE-2 through CSE-5) use only $O(nm)$ clauses. Similarly, CSE-6 encodes precedence (different stations) with $(m-1)$ binary clauses per edge versus SAT-3's $\binom{m}{2}$ clauses, achieving factor-$m$ reduction. Each CSE-6 clause contains exactly 2 literals (enabling efficient unit propagation) versus up to $m$ literals in SAT-3.

For same-station precedence (CSE-13), the cumulative time variable $T_{j,t}$ enables a linear reformulation. SAT-4 enumerates all pairs $(t_1, t_2)$ where $t_1 > t_2$, yielding $O(c^2)$ clauses per edge-workstation pair. In contrast, CSE-13 iterates once over start times $t$ for task $i$, using $T_{j,t'}$ to compactly represent ``task $j$ started by time $t'$'', reducing complexity from $O(c^2)$ to $O(c)$ per edge-workstation pair.

\subsection{Extended Precedence Graph Preprocessing}
\label{subsec:extended_precedence}

Beyond the compact encoding, we apply extended precedence preprocessing to exploit transitive closure of the precedence DAG. Given $G = (V, E)$ with direct precedence edges, we compute $G^* = (V, E^*)$ where $E^* = \{(i, j) : \exists \text{ directed path from } i \text{ to } j \text{ in } G\}$. The algorithm uses depth-first search with memoization in $O(n \cdot (n + |E|))$ time (detailed in Appendix~\ref{appendix:extended_precedence}).

We then apply all precedence constraints (CSE-6, CSE-13) to $E^*$ instead of $E$. While constraints for $(i,j) \in E^* \setminus E$ are logically redundant, their explicit encoding provides immediate propagation when assigning tasks, direct conflict detection, and tighter preprocessing bounds for $\text{First}(i)$ and $\text{Last}(i)$. This technique was demonstrated effective for SMT-based scheduling \citep{bofill2017smt}, where transitive closure preprocessing significantly reduces search space for problems with dense precedence structures.

% \paragraph{Trade-off Analysis.}
% While CSE achieves $O(m)$ clauses per edge compared to ORG's $O(m^2)$, CSE applies constraints to the transitive closure $E^*$ rather than the original edge set $E$. In the worst case (e.g., a chain graph $1 \to 2 \to \cdots \to n$), we have $|E| = n-1$ but $|E^*| = n(n-1)/2 = O(n^2)$. For such graphs, CSE generates $O(|E^*| \cdot m) = O(n^2 m)$ precedence clauses while ORG generates $O(|E| \cdot m^2) = O(nm^2)$. When $n > m$ (typical in SALBP where tasks outnumber workstations), CSE may produce more precedence clauses than ORG for chain-like structures. However, CSE's primary gains come from the at-most-one (AMO) constraints for task assignment ($O(nm)$ vs $O(nm^2)$) and scheduling ($O(nc)$ vs $O(nc^2)$), which dominate the total clause count for practical instances. Furthermore, assembly line precedence graphs are typically sparse with bounded transitive closure density, and the explicit transitive constraints improve solver propagation efficiency, as demonstrated empirically in Section~\ref{sec:results}.

\subsection{Optimization Strategies}
\label{subsec:optimization_strategies}

We propose four algorithmic variants for peak power minimization, each building upon our compact precedence encoding (Section~\ref{subsec:compact_encoding}). Let $\Phi_{base}$ denote the compact constraints (CSE-1 through CSE-16) common to all variants.

\subsubsection{Enhanced Upper Bound Initialization}
\label{subsubsec:tight_ub}

Traditional analytical estimates like $UB = \sum_{w_i \in W_m} w_i$ often overestimate achievable peaks. We compute SAT-derived bounds through a multi-solve initialization phase (Algorithm~\ref{alg:ub_init}) that iteratively solves $\Phi_{base}$, extracts peak configurations, and adds blocking clauses forcing alternative solutions. The number of initialization iterations is a tunable parameter balancing bound tightness against computational overhead; in our experiments, we use 10 iterations as a practical trade-off. The minimum observed peak becomes $UB_{tight}$, used by MaxSAT and Incremental SAT variants (but not Clause-Based or PB-Constraint variants, which start from the first feasible solution).

\subsubsection{Alternative Optimization Variants}
\label{subsubsec:alternative_variants}

We implement three baseline variants for comparative evaluation.

\emph{Clause-Based Iterative Minimization (CSE\_CB)} starts from the first feasible solution and maintains persistent solver state across iterations, progressively excluding high-peak configurations through blocking clauses. Each blocking clause $\bigvee_{i \in C} \overline{A_{i,t}}$ forbids task configurations causing peaks $\geq W_{current}$, preserving learned clauses for CDCL reuse.

\emph{PB-Constraint Iterative Optimization (CSE\_PB)} also starts from the first feasible solution, then employs pseudo-Boolean constraints $\sum_{i=1}^n w_i \cdot A_{i,t} \leq W_{best} - 1$ to enforce strict peak reduction. This variant reconstructs the solver at each iteration with tightened bounds encoded via totalizer networks.

\emph{MaxSAT Formulation (CSE\_MaxHS, CSE\_Eval)} first obtains $UB_{tight}$ via the initialization phase, then reformulates peak minimization as weighted partial MaxSAT, combining our compact encoding with the binary peak representation of \citet{Zheng2024MaxSAT}. The peak is encoded using $\lceil \log_2 UB_{tight} \rceil$ binary variables with soft clauses $(\neg binU_j, 2^j)$ for single-shot optimization.

Detailed algorithms for these variants are provided in Appendix~\ref{appendix:algorithms}.

\subsubsection{Incremental SAT with Peak Indicators}
\label{subsubsec:incremental_sat}

Our primary contribution is an incremental SAT approach that preserves solver state through peak indicator variables $\{U_j\}_{j=LB+1}^{UB_{tight}-1}$, adapted from the inaugural MaxSAT model of \citet{Zheng2024MaxSAT}. These variables satisfy $U_j = 1$ iff the peak power is $\geq j$. After obtaining $UB_{tight}$ from the initialization phase, this approach enables peak refinement via simple unit clause assertions ($\neg U_j$), avoiding costly solver reconstruction while preserving learned clauses.

\begin{algorithm}[!htb]
\caption{Incremental SAT with Peak Refinement}
\label{alg:incremental_sat}
\begin{algorithmic}[1]
\Require Instance parameters, timeout limit
\Ensure Optimal solution with minimum peak power
\State $(W_{best}, \phi_{best}) \leftarrow$ \textsc{InitializeBounds}($\Phi_{base}$) \Comment{Initial solve phase}
\State $LB \leftarrow \max_{i} w_i$
\State Initialize incremental solver $\mathcal{S}$ with $\Phi_{base}$
\State Generate peak indicators $U_j$ for $j \in \{LB+1, \ldots, W_{best}-1\}$
\State Add ordering: $\neg U_j \vee U_{j-1}$ for $j \in \{LB+2, \ldots, W_{best}-1\}$ to $\mathcal{S}$
\For{each time slot $t \in T$}
    \State $\Phi_{PB} \leftarrow \textsc{EncodePBtoCNF}(\{\neg U_j\}, \{A_{i,t}\}, \{1\}, \{w_i\}, W_{best})$
    \State Add $\Phi_{PB}$ to $\mathcal{S}$ \Comment{$\sum_j \neg U_j + \sum_i w_i A_{i,t} \leq W_{best}$}
\EndFor
\While{time limit not exceeded}
    \State $\phi \leftarrow \textsc{Solve}(\mathcal{S})$
    \If{$\phi = $ UNSAT} \Return $\phi_{best}, W_{best}$
    \EndIf
    \State $W_{current} \leftarrow \max_{t} \sum_{i: A_{i,t}=\text{true}} w_i$
    \If{$W_{current} < W_{best}$} $\phi_{best} \leftarrow \phi$; $W_{best} \leftarrow W_{current}$
    \EndIf
    \State Add unit clause $\neg U_{W_{current}-1}$ to $\mathcal{S}$ \Comment{Forbid peak $\geq W_{current}$}
\EndWhile
\State \Return $\phi_{best}, W_{best}$
\end{algorithmic}
\end{algorithm}

The key insight is that using $O(UB_{tight} - LB)$ peak indicator variables $U_j$ enables natural incrementality: the number of true $U_j$ variables determines the allowed peak upper bound via the PB constraint $\sum_j \neg U_j + \sum_i w_i A_{i,t} \leq UB_{tight}$. When a solution with peak $W_{current}$ is found, adding $\neg U_{W_{current}-1}$ forces subsequent solutions to have peak $< W_{current}$, since at most $W_{current} - LB - 2$ indicator variables can be true, yielding peak $\leq W_{current} - 1$. Ordering constraints $\neg U_j \vee U_{j-1}$ maintain monotonicity: if $U_j$ is true then $U_{j-1}$ must also be true.

This approach combines the advantages of both iterative SAT (preserving learned clauses) and MaxSAT (explicit peak bounds) while avoiding their respective drawbacks: unlike Clause-Based which generates exponentially many blocking clauses, Incremental SAT uses a fixed set of peak indicators; unlike MaxSAT which requires external solvers, it leverages the highly optimized CaDiCaL incremental interface. Experimental results (Section~\ref{sec:results}) confirm that CSE\_INC achieves the best overall performance.

\section{Experimental Setup}\label{sec:experiments}

\subsection{Benchmark Instances}

Our experimental dataset comprises 89 SALBP-3PM instances derived from 13 classical SALBP benchmark families available from the assembly line balancing repository\footnote{\url{https://assembly-line-balancing.de/salbp/benchmark-data-sets-1993/}}. The instances range from small problems (7 tasks, 2 workstations) to large-scale problems (89 tasks, 49 workstations), providing comprehensive coverage of problem complexity.

Following the established benchmark protocol \citep{Gianessi2019SALB3PM,Py2024POS,Zheng2024MaxSAT}, each SALBP-3PM instance is constructed by taking precedence constraints $\mathcal{P}$ and task processing times $\{d_i\}$ directly from the original SALBP datasets, setting cycle time $c$ to the optimal cycle time from the corresponding SALBP-1 solution, fixing the number of workstations $m$ to the optimal value $m^*$ from SALBP-1, and randomly generating power consumption $\{w_i\}$ for each task following the standard protocol. Table~\ref{tab:benchmarks} summarizes the 89 benchmark instances from 13 families, grouped by problem size.

\begin{table}[htbp]
\caption{Benchmark instances grouped by problem size.}
\label{tab:benchmarks}
\small
\centering
\begin{tabular}{@{}llcc@{}}
\toprule
\textbf{Size} & \textbf{Family} & \textbf{\#Tasks} & \textbf{\#Instances} \\
\midrule
\multirow{5}{*}{Small} 
& MERTENS & 7 & 6 \\
& BOWMAN & 8 & 1 \\
& JAESCHKE & 9 & 5 \\
& JACKSON & 11 & 6 \\
& MANSOOR & 11 & 3 \\
\midrule
\multirow{5}{*}{Medium} 
& MITCHELL & 21 & 6 \\
& ROSZIEG & 25 & 6 \\
& HESKIA & 28 & 6 \\
& BUXEY & 29 & 7 \\
& SAWYER & 30 & 9 \\
\midrule
\multirow{3}{*}{Large} 
& GUNTHER & 35 & 7 \\
& WARNECKE & 58 & 16 \\
& LUTZ2 & 89 & 11 \\
\bottomrule
\end{tabular}
\end{table}

\subsection{Experimental Environment and Metrics}

All experiments are conducted on Google Cloud Platform using c4-highcpu-8 instances (8 vCPUs, 16 GB Memory) running Ubuntu 20.04 LTS. Each instance is solved with a timeout limit of 3600 seconds. For reproducibility, we use consistent random seeds across all solvers. Our solver suite includes CaDiCaL 1.9.5 for iterative and incremental SAT variants, EvalMaxSAT and MaxHS for weighted MaxSAT formulations, PySAT library for pseudo-boolean to CNF translation using the Binary Merger encoding \citep{manthey2014more}, and IBM CPLEX 22.1.1 as ILP baseline using the established ILP formulation for SALBP-3PM \citep{Gianessi2019SALB3PM}. 
% All source code, benchmark datasets, and detailed experimental results are publicly available at \url{https://github.com/salbp-3pm/cse}.

Our evaluation employs two primary performance metrics: (1) the number of instances solved to optimality within the 3600-second timeout, which directly measures practical effectiveness, and (2) total solving time for solved instances to compare computational efficiency. All reported solving times include the complete algorithm execution, including any initialization phases (such as bound estimation) as part of the total time measurement.

Table~\ref{tab:configurations} summarizes all nine solver configurations evaluated in our experiments. The key distinction between ORG\_ and CSE\_ methods lies in the constraint encoding: ORG\_ methods use baseline encodings from prior work \citep{Py2024POS,Zheng2024MaxSAT}, while CSE\_ methods employ our proposed Compact SAT Encoding requiring only $O(m)$ clauses per transitive precedence edge (versus $O(m^2)$ for baseline approaches).

\begin{table}[htbp]
\caption{Solver configurations compared in experimental evaluation.}
\label{tab:configurations}
\small
\centering
\begin{tabular}{@{}lll@{}}
\toprule
\textbf{Category} & \textbf{Configuration} & \textbf{Description} \\
\midrule
ILP Baseline 
& CPLEX & Commercial ILP solver \citep{Gianessi2019SALB3PM} \\
\midrule
\multirow{3}{*}{Prior Work} 
& ORG\_CB & Iterative SAT with clause blocking \citep{Py2024POS} \\
& ORG\_MaxHS & MaxSAT (MaxHS solver) \citep{Zheng2024MaxSAT} \\
& ORG\_Eval & MaxSAT (EvalMaxSAT solver) \citep{Zheng2024MaxSAT} \\
\midrule
\multirow{5}{*}{Proposed (CSE)} 
& CSE\_CB & Iterative SAT with clause blocking (Appendix~\ref{appendix:algorithms}) \\
& CSE\_PB & Iterative SAT with PB constraints (Appendix~\ref{appendix:algorithms}) \\
& CSE\_MaxHS & MaxSAT (MaxHS solver) (Appendix~\ref{appendix:algorithms}) \\
& CSE\_Eval & MaxSAT (EvalMaxSAT solver) (Appendix~\ref{appendix:algorithms}) \\
& CSE\_INC & Incremental SAT refinement (Sec.~\ref{subsubsec:incremental_sat}) \\
\bottomrule
\end{tabular}
\end{table}

\section{Experimental Results}\label{sec:results}

We present our experimental evaluation in three parts to comprehensively assess the performance of our proposed Compact SAT Encoding (CSE) methods: (1) overall performance comparison across all benchmark families, (2) detailed analysis of hard instances requiring over 100 seconds, and (3) extended timeout experiments on extreme-hard instances that were previously unsolvable.

\subsection{Overall Performance Comparison}\label{subsec:overall_perf}

Table~\ref{tab:overall_results} summarizes the performance of all configurations across 13 benchmark families comprising 89 instances total, with a 3600-second timeout. Each cell shows the number of instances solved to optimality along with the cumulative solving time (in seconds) for those solved instances.

\begin{sidewaystable}[htbp]
\caption{Performance comparison across benchmark families (\#Opt / Time in seconds). Timeout: 3600s.}
\label{tab:overall_results}
\small
\centering
\newcommand{\cellbb}[2]{\begin{tabular}[c]{@{}c@{}}\textbf{#1}\\\textbf{#2}\end{tabular}}
\newcommand{\cellbt}[2]{\begin{tabular}[c]{@{}c@{}}#1\\\textbf{#2}\end{tabular}}
\newcommand{\celltb}[2]{\begin{tabular}[c]{@{}c@{}}\textbf{#1}\\#2\end{tabular}}
\newcommand{\cellnn}[2]{\begin{tabular}[c]{@{}c@{}}#1\\#2\end{tabular}}
\newcommand{\cellna}{\begin{tabular}[c]{@{}c@{}}0\\---\end{tabular}}
\newcommand{\thinrule}{\specialrule{0.1pt}{0.25pt}{0.25pt}}
\begin{tabular}{@{}lccccccccc@{}}
\toprule
\multirow{2}{*}{Family} & \multicolumn{1}{c}{ILP} & \multicolumn{3}{c}{Prior Work (ORG\_)} & \multicolumn{5}{c}{Proposed (CSE\_)} \\
\cmidrule(lr){2-2} \cmidrule(lr){3-5} \cmidrule(lr){6-10}
& CPLEX & CB & MaxHS & Eval & CB & PB & MaxHS & Eval & INC \\
\midrule
BOWMAN (1)     & \celltb{1}{1.48} & \cellbb{1}{0.01} & \celltb{1}{0.06} & \celltb{1}{0.04} & \cellbb{1}{0.01} & \cellbb{1}{0.01} & \celltb{1}{0.03} & \celltb{1}{0.02} & \cellbb{1}{0.01} \\[4pt]
\thinrule
JACKSON (6)    & \celltb{6}{6.47} & \celltb{6}{0.14} & \celltb{6}{0.55} & \celltb{6}{0.77} & \cellbb{6}{0.11} & \celltb{6}{0.25} & \celltb{6}{0.28} & \celltb{6}{0.21} & \celltb{6}{0.53} \\[4pt]
\thinrule
JAESCHKE (5)   & \celltb{5}{1.84} & \cellbb{5}{0.04} & \celltb{5}{0.19} & \celltb{5}{0.20} & \cellbb{5}{0.04} & \celltb{5}{0.06} & \celltb{5}{0.09} & \celltb{5}{0.06} & \celltb{5}{0.07} \\[4pt]
\thinrule
MANSOOR (3)    & \celltb{3}{20.38} & \celltb{3}{0.24} & \celltb{3}{1.60} & \celltb{3}{3.19} & \celltb{3}{0.13} & \celltb{3}{0.14} & \cellbb{3}{0.09} & \cellbb{3}{0.09} & \celltb{3}{0.15} \\[4pt]
\thinrule
MERTENS (6)    & \celltb{6}{1.32} & \celltb{6}{0.02} & \celltb{6}{0.26} & \celltb{6}{0.81} & \celltb{6}{0.03} & \celltb{6}{0.02} & \cellbb{6}{0.01} & \cellbb{6}{0.01} & \celltb{6}{0.02} \\[4pt]
\thinrule
MITCHELL (6)   & \celltb{6}{2330.55} & \celltb{6}{4.89} & \celltb{6}{12.53} & \celltb{6}{13.00} & \cellbb{6}{2.33} & \celltb{6}{5.94} & \celltb{6}{4.52} & \celltb{6}{4.83} & \celltb{6}{5.64} \\[4pt]
\thinrule
ROSZIEG (6)    & \celltb{6}{10760.06} & \celltb{6}{120.17} & \celltb{6}{655.70} & \cellnn{5}{32.22} & \celltb{6}{110.11} & \celltb{6}{63.85} & \celltb{6}{231.57} & \celltb{6}{142.02} & \cellbb{6}{60.64} \\[4pt]
\thinrule
HESKIA (6)     & \cellna & \cellnn{1}{1259.62} & \cellnn{1}{1582.03} & \cellnn{3}{4237.13} & \celltb{5}{7559.03} & \celltb{5}{2256.99} & \celltb{5}{2301.49} & \celltb{5}{1660.09} & \cellbb{5}{1488.82} \\[4pt]
\thinrule
BUXEY (7)      & \cellnn{1}{304.26} & \cellnn{3}{624.13} & \cellnn{4}{2578.25} & \cellnn{4}{1081.39} & \cellnn{3}{1173.06} & \celltb{5}{3166.54} & \cellnn{4}{2659.32} & \celltb{5}{4444.36} & \cellbb{5}{1876.97} \\[4pt]
\thinrule
SAWYER (9)     & \cellnn{1}{1401.18} & \cellnn{4}{3236.74} & \celltb{5}{1326.26} & \cellnn{4}{1067.33} & \cellnn{4}{4486.11} & \celltb{5}{880.86} & \celltb{5}{1649.51} & \celltb{5}{1131.58} & \cellbb{5}{653.57} \\[4pt]
\thinrule
GUNTHER (7)    & \cellna & \celltb{5}{3545.38} & \celltb{5}{1329.31} & \celltb{5}{760.66} & \celltb{5}{3518.84} & \celltb{5}{351.81} & \celltb{5}{1054.93} & \celltb{5}{578.96} & \cellbb{5}{334.49} \\[4pt]
\thinrule
WARNECKE (16)   & \cellna & \cellna & \cellna & \cellna & \cellna & \cellna & \cellna & \cellna & \cellbb{1}{2040.49} \\[4pt]
\thinrule
LUTZ2 (11)      & \cellna & \celltb{2}{439.58} & \cellbb{2}{23.71} & \celltb{2}{143.17} & \celltb{2}{106.89} & \celltb{2}{767.21} & \cellbb{2}{23.71} & \celltb{2}{57.91} & \celltb{2}{93.19} \\
\midrule
\begin{tabular}[c]{@{}l@{}}\textbf{Total (89)}\end{tabular} & \cellnn{35}{14827.54} & \cellnn{48}{9230.97} & \cellnn{50}{7548.99} & \cellnn{50}{7339.91} & \cellnn{52}{16956.68} & \cellnn{55}{7493.66} & \cellnn{54}{7925.55} & \cellnn{55}{8020.15} & \cellbb{56}{6554.71} \\
\bottomrule
\end{tabular}
\end{sidewaystable}

Our proposed CSE configurations demonstrate significant performance improvements across all evaluation metrics compared to existing approaches. As shown in Table~\ref{tab:overall_results}, CSE\_INC achieves the best overall performance with 56 instances solved (62.9\% of total), compared to 50 for the best prior method ORG\_Eval (56.2\%), representing a 12\% improvement in solved instances that validates the effectiveness of combining Compact SAT Encoding with incremental SAT optimization. To isolate the encoding contribution from algorithmic factors, we compare CSE\_CB (52 solved) versus ORG\_CB (48 solved)—both employing iterative SAT with identical solving strategies but different encodings—which demonstrates that Compact SAT Encoding alone enables solving 4 additional instances, directly validating the theoretical compactness claims from Table~\ref{tab:complexity_comparison}.

Beyond solution coverage, the compact encoding approach exhibits superior computational efficiency. CSE\_INC achieves the lowest total solving time (6554.71s) among all configurations, representing a 10.7\% reduction compared to the best prior method ORG\_Eval (7339.91s). This dual advantage—higher solution coverage and faster solving on successfully solved instances—confirms that compact encoding provides fundamental improvements rather than trading off coverage against speed. Among the CSE variants, CSE\_Eval demonstrates strong performance with 55 instances solved, often matching CSE\_INC on smaller benchmark families (BOWMAN, JACKSON, JAESCHKE, MERTENS) while CSE\_INC shows advantages on larger, more complex instances (HESKIA, SAWYER, WARNECKE), suggesting that problem characteristics influence the optimal algorithmic choice within the compact encoding framework.

Family-specific performance analysis further reveals the transformative impact of compact encoding on challenging benchmarks. On the WARNECKE family (16 instances), only CSE methods achieve any solutions, with CSE\_Eval and CSE\_INC each solving 1 instance while all prior methods fail completely. On the HESKIA family (6 instances), CSE methods solve 5 instances compared to only 3 for ORG\_Eval, demonstrating substantial improvements on complex problems with intricate precedence structures. In contrast, CPLEX solves only 35 instances (39.3\%), significantly underperforming all SAT-based methods, confirming that specialized SAT encodings are essential for SALBP-3PM as commercial ILP solvers struggle with the problem's combinatorial structure.

\subsection{Encoding Size Analysis}\label{subsec:encoding_size}

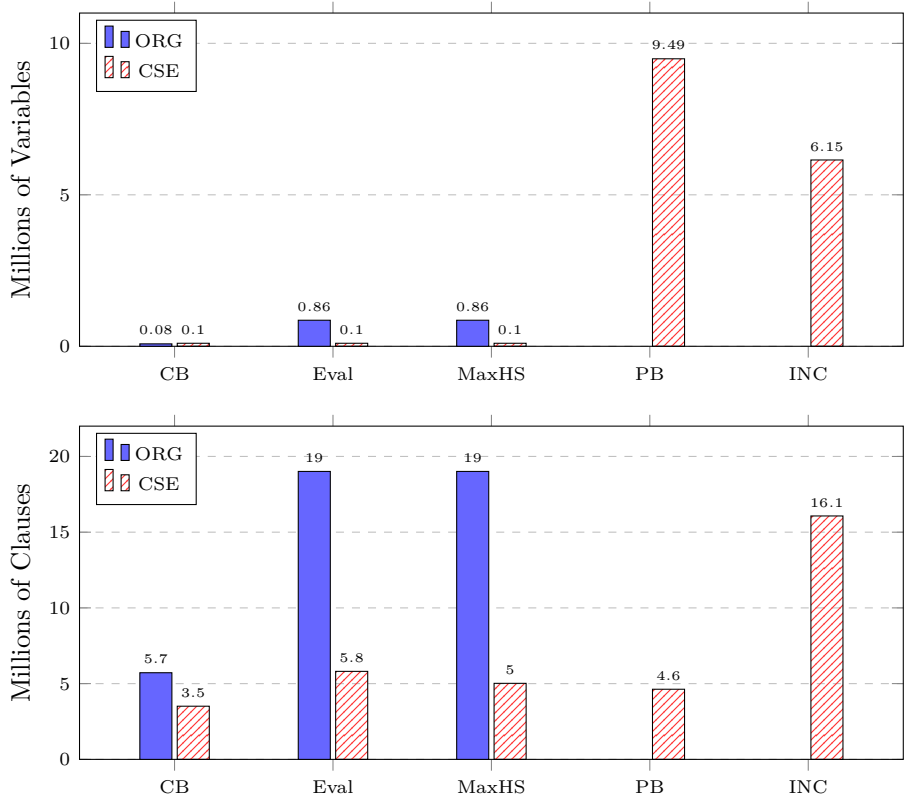
\begin{figure}[htbp]
\centering
% Variables chart (top)
\begin{tikzpicture}
\begin{axis}[
    ybar,
    width=0.95\textwidth,
    height=6cm,
    ylabel={Millions of Variables},
    symbolic x coords={CB, Eval, MaxHS, PB, INC},
    xtick={CB, Eval, MaxHS, PB, INC},
    xticklabel style={font=\footnotesize},
    yticklabel style={font=\footnotesize},
    ymajorgrids=true,
    grid style=dashed,
    bar width=12pt,
    ymin=0,
    ymax=11,
    nodes near coords,
    nodes near coords style={font=\tiny, above},
    every node near coord/.append style={/pgf/number format/.cd, fixed, precision=2},
    legend style={at={(0.02,0.98)}, anchor=north west, font=\footnotesize},
    enlarge x limits=0.15,
]
% ORG methods (only CB, Eval, MaxHS)
\addplot[fill=blue!60] coordinates {
    (CB, 0.08) (Eval, 0.86) (MaxHS, 0.86)
};
% CSE methods (all five)
\addplot[pattern=north east lines, pattern color=red!80] coordinates {
    (CB, 0.10) (Eval, 0.10) (MaxHS, 0.10) (PB, 9.49) (INC, 6.15)
};
\legend{ORG, CSE}
\end{axis}
\end{tikzpicture}

\vspace{0.3cm}

% Clauses chart (bottom)
\begin{tikzpicture}
\begin{axis}[
    ybar,
    width=0.95\textwidth,
    height=6cm,
    ylabel={Millions of Clauses},
    symbolic x coords={CB, Eval, MaxHS, PB, INC},
    xtick={CB, Eval, MaxHS, PB, INC},
    xticklabel style={font=\footnotesize},
    yticklabel style={font=\footnotesize},
    ymajorgrids=true,
    grid style=dashed,
    bar width=12pt,
    ymin=0,
    ymax=22,
    nodes near coords,
    nodes near coords style={font=\tiny, above},
    every node near coord/.append style={/pgf/number format/.cd, fixed, precision=1},
    legend style={at={(0.02,0.98)}, anchor=north west, font=\footnotesize},
    enlarge x limits=0.15,
]
% ORG methods (only CB, Eval, MaxHS)
\addplot[fill=blue!60] coordinates {
    (CB, 5.72) (Eval, 19.01) (MaxHS, 19.01)
};
% CSE methods (all five)
\addplot[pattern=north east lines, pattern color=red!80] coordinates {
    (CB, 3.51) (Eval, 5.81) (MaxHS, 5.02) (PB, 4.63) (INC, 16.07)
};
\legend{ORG, CSE}
\end{axis}
\end{tikzpicture}
\caption{Encoding size comparison across 45 commonly-solved instances. Top: total variables (millions); Bottom: total clauses (millions). Solid blue bars represent ORG (baseline encoding); hatched red bars represent CSE (compact encoding)}
\label{fig:encoding_size}
\end{figure}

To validate the theoretical complexity reduction from Table~\ref{tab:complexity_comparison}, we analyze the actual encoding sizes across 45 instances solved by all SAT/MaxSAT methods. Figure~\ref{fig:encoding_size} compares the total number of clauses and variables for each configuration.

The encoding size analysis validates the theoretical complexity established in Tables~\ref{tab:variable_comparison} and~\ref{tab:complexity_comparison}. A critical insight emerges: clause count, not variable count, determines solver efficiency. As shown in Figure~\ref{fig:encoding_size}, CSE\_CB achieves the smallest clause count (3.51M), representing an 81.5\% reduction compared to ORG\_Eval/ORG\_MaxHS (19.01M). This dramatic reduction directly validates the theoretical $O(nm^2) \to O(nm)$ complexity improvement for AMO constraints.

The variable comparison reveals a nuanced picture consistent with Table~\ref{tab:variable_comparison}. For base SAT configurations, CSE\_CB uses 0.10M variables versus ORG\_CB's 0.08M—a modest 25\% increase reflecting the $O(nm + nc)$ auxiliary reach variables ($R_{i,k}$, $T_{i,t}$). Both ORG and CSE MaxSAT methods employ binary peak encoding with $O(\log_2 UB)$ variables. However, CSE methods benefit from the tighter bound $UB_{tight}$ obtained through the multi-solve initialization phase (Section~\ref{subsubsec:tight_ub}), reducing both peak indicator variables and objective-related clauses. This explains why ORG\_Eval/ORG\_MaxHS require 0.86M variables (using analytical $UB = \sum_{w_i \in W_m} w_i$) while CSE\_Eval/CSE\_MaxHS use only 0.10M variables with the same binary encoding structure.

The outliers CSE\_INC (16.07M clauses, 6.15M variables) and CSE\_PB (4.63M clauses, 9.49M variables) reflect algorithm-specific overheads. CSE\_INC accumulates blocking clauses and creates peak indicator variables $U_j$ for $j \in \{LB+1, \ldots, UB_{tight}\}$ across iterations. CSE\_PB introduces auxiliary variables for totalizer-based pseudo-Boolean to CNF translation, where encoding $\sum_i w_i A_{i,t} \leq W$ with diverse coefficients $w_i$ requires $O(n \cdot UB)$ auxiliary variables.

The apparent paradox that CSE\_INC outperforms other CSE variants despite having more clauses is explained by three critical factors. First, \emph{solver state preservation} plays a key role: unlike CSE\_CB which reconstructs the solver from scratch after each blocking clause, CSE\_INC maintains persistent solver state through CaDiCaL's incremental interface, preserving learned conflict clauses that accelerate subsequent iterations. Second, \emph{structured clause growth} contributes to efficiency: the additional clauses in CSE\_INC arise from peak indicator encoding with well-structured monotonic relationships that enable efficient unit propagation, unlike the arbitrary blocking clauses in iterative methods. Third, \emph{reduced iteration overhead} provides further advantage: while Figure~\ref{fig:encoding_size} reports final encoding sizes after all iterations, CSE\_INC's incremental nature means each iteration adds only a single unit clause ($\neg U_j$), whereas CSE\_CB and CSE\_MaxHS incur substantial per-iteration overhead from blocking clause generation or solver reconstruction. The combination of learned clause preservation and structured incremental refinement explains why CSE\_INC achieves both the highest solution count and lowest total solving time despite its larger final encoding.

\subsection{Performance on Hard Benchmarks}\label{subsec:hard_benchmarks}

\begin{figure}[htbp]
\centering
\begin{tikzpicture}
\begin{axis}[
    width=\textwidth,
    height=7cm,
    ylabel={Solving Time (seconds, log scale)},
    symbolic x coords={B\_29-11-33, B\_29-12-30, G\_35-9-61, H\_28-3-342, H\_28-4-256, H\_28-5-205, H\_28-5-216, H\_28-8-138, S\_30-12-30, W\_58-25-65},
    xtick=data,
    xticklabel style={rotate=45, anchor=east, font=\footnotesize},
    yticklabel style={font=\footnotesize},
    legend style={at={(0.5,-0.35)}, anchor=north, legend columns=3, font=\tiny},
    ymajorgrids=true,
    grid style=dashed,
    enlarge x limits=0.05,
    ymode=log,
    ymin=50,
    ymax=5000,
    log origin=infty,
    line width=1pt,
]
% CPLEX - all timeout (3600s)
\addplot[color=gray, mark=square*, mark size=1.8pt, densely dashed] coordinates {
    (B\_29-11-33,3600) (B\_29-12-30,3600) (G\_35-9-61,3600) (H\_28-3-342,3600) (H\_28-4-256,3600) (H\_28-5-205,3600) (H\_28-5-216,3600) (H\_28-8-138,3600) (S\_30-12-30,3600) (W\_58-25-65,3600)
};
% ORG_CB - 2 solved, 8 timeout
\addplot[color=red!70, mark=triangle*, mark size=2pt, dotted] coordinates {
    (B\_29-11-33,3600) (B\_29-12-30,3600) (G\_35-9-61,3532.66) (H\_28-3-342,3600) (H\_28-4-256,3600) (H\_28-5-205,3600) (H\_28-5-216,3600) (H\_28-8-138,1259.62) (S\_30-12-30,3600) (W\_58-25-65,3600)
};
% ORG_MaxHS - 3 solved, 7 timeout
\addplot[color=orange!80, mark=diamond*, mark size=2pt, dashdotted] coordinates {
    (B\_29-11-33,2409.04) (B\_29-12-30,3600) (G\_35-9-61,1307.72) (H\_28-3-342,3600) (H\_28-4-256,3600) (H\_28-5-205,3600) (H\_28-5-216,3600) (H\_28-8-138,1582.03) (S\_30-12-30,995.57) (W\_58-25-65,3600)
};
% ORG_Eval - 5 solved, 5 timeout
\addplot[color=olive, mark=star, mark size=2pt, densely dotted] coordinates {
    (B\_29-11-33,934.06) (B\_29-12-30,3600) (G\_35-9-61,731.08) (H\_28-3-342,3600) (H\_28-4-256,3600) (H\_28-5-205,1231.83) (H\_28-5-216,2461.68) (H\_28-8-138,543.62) (S\_30-12-30,850.04) (W\_58-25-65,3600)
};
% CSE_CB - 6 solved, 4 timeout
\addplot[color=cyan!70, mark=pentagon*, mark size=2pt, dashed] coordinates {
    (B\_29-11-33,3600) (B\_29-12-30,3600) (G\_35-9-61,3480.56) (H\_28-3-342,740.89) (H\_28-4-256,1910.24) (H\_28-5-205,828.80) (H\_28-5-216,1739.17) (H\_28-8-138,2339.94) (S\_30-12-30,3600) (W\_58-25-65,3600)
};
% CSE_PB - 8 solved, 2 timeout
\addplot[color=blue!70, mark=square*, mark size=1.8pt] coordinates {
    (B\_29-11-33,536.80) (B\_29-12-30,2532.26) (G\_35-9-61,309.73) (H\_28-3-342,400.27) (H\_28-4-256,818.52) (H\_28-5-205,202.55) (H\_28-5-216,436.38) (H\_28-8-138,399.26) (S\_30-12-30,635.19) (W\_58-25-65,3600)
};
% CSE_MaxHS - 7 solved, 3 timeout
\addplot[color=purple!70, mark=triangle*, mark size=2pt] coordinates {
    (B\_29-11-33,2582.05) (B\_29-12-30,3600) (G\_35-9-61,1041.61) (H\_28-3-342,258.27) (H\_28-4-256,420.15) (H\_28-5-205,215.88) (H\_28-5-216,893.59) (H\_28-8-138,513.60) (S\_30-12-30,1328.44) (W\_58-25-65,3600)
};
% CSE_Eval - 8 solved, 2 timeout
\addplot[color=green!60!black, mark=diamond*, mark size=2pt] coordinates {
    (B\_29-11-33,923.08) (B\_29-12-30,3440.18) (G\_35-9-61,568.61) (H\_28-3-342,243.26) (H\_28-4-256,314.07) (H\_28-5-205,122.64) (H\_28-5-216,491.06) (H\_28-8-138,489.05) (S\_30-12-30,963.83) (W\_58-25-65,3600)
};
% CSE_INC (best) - 10/10 solved
\addplot[color=red!80!black, mark=*, mark size=2.5pt, line width=1.5pt] coordinates {
    (B\_29-11-33,549.21) (B\_29-12-30,1272.76) (G\_35-9-61,300.80) (H\_28-3-342,181.61) (H\_28-4-256,468.82) (H\_28-5-205,150.19) (H\_28-5-216,463.43) (H\_28-8-138,224.77) (S\_30-12-30,536.31) (W\_58-25-65,2040.49)
};

\legend{CPLEX, ORG\_CB, ORG\_MaxHS, ORG\_Eval, CSE\_CB, CSE\_PB, CSE\_MaxHS, CSE\_Eval, CSE\_INC}
\end{axis}
\end{tikzpicture}
\caption{Solving times (seconds, log scale) for 10 hard instances. Timeout (3600s) shown for unsolved cases. Instance abbreviations: B=BUXEY, G=GUNTHER, H=HESKIA, S=SAWYER, W=WARNECKE}
\label{fig:hard_benchmarks}
\end{figure}
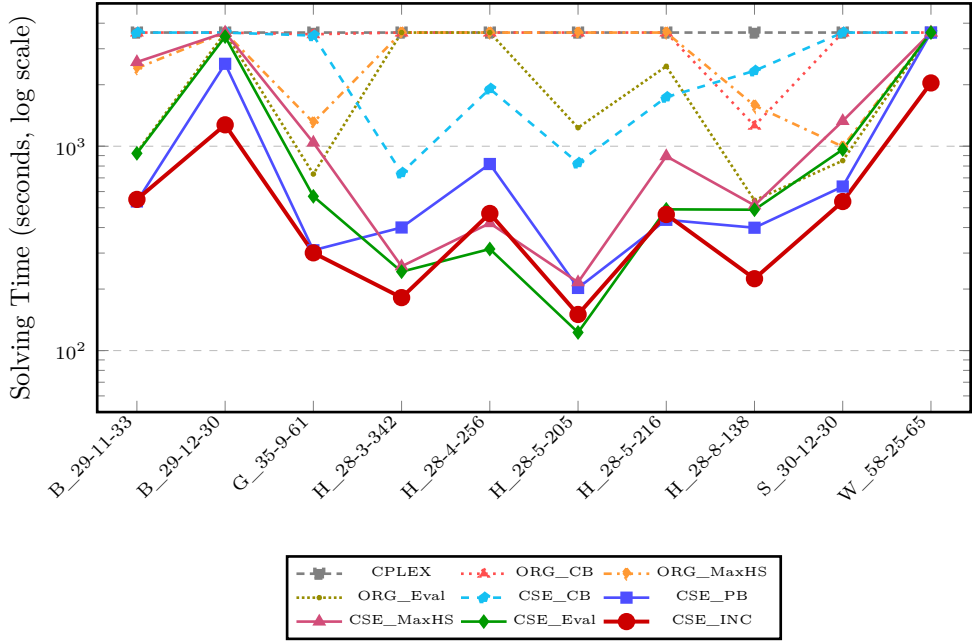

To assess scalability on challenging instances, we focus on benchmarks where solving difficulty is substantial. Table~\ref{tab:hard_benchmarks} provides detailed solving times for 10 hard instances requiring over 100 seconds for at least one top-performing configuration, isolating cases where solver efficiency differences become most pronounced. Figure~\ref{fig:hard_benchmarks} visualizes these results.

\begin{sidewaystable}[htbp]
\caption{Performance on hard benchmarks (instances requiring more than 100 seconds for at least one top configuration). Times in seconds; --- indicates timeout at 3600s.}
\label{tab:hard_benchmarks}
\small
\centering
\begin{tabular}{@{}lcccccccccc@{}}
\toprule
\multirow{2}{*}{Instance} & \multicolumn{1}{c}{ILP} & \multicolumn{3}{c}{Prior Work (ORG\_)} & \multicolumn{5}{c}{Proposed (CSE\_)} \\
\cmidrule(lr){2-2} \cmidrule(lr){3-5} \cmidrule(lr){6-10}
& \multicolumn{1}{c}{CPLEX} & \multicolumn{1}{c}{CB} & \multicolumn{1}{c}{MaxHS} & \multicolumn{1}{c}{Eval} & \multicolumn{1}{c}{CB} & \multicolumn{1}{c}{PB} & \multicolumn{1}{c}{MaxHS} & \multicolumn{1}{c}{Eval} & \multicolumn{1}{c}{INC} \\
\midrule
BUXEY\_29-11-33   & --- & --- & 2409.04 & 934.06 & --- & \textbf{536.80} & 2582.05 & 923.08 & 549.21 \\
BUXEY\_29-12-30   & --- & --- & --- & --- & --- & 2532.26 & --- & 3440.18 & \textbf{1272.76} \\
GUNTHER\_35-9-61  & --- & 3532.66 & 1307.72 & 731.08 & 3480.56 & 309.73 & 1041.61 & 568.61 & \textbf{300.80} \\
HESKIA\_28-3-342  & --- & --- & --- & --- & 740.89 & 400.27 & 258.27 & 243.26 & \textbf{181.61} \\
HESKIA\_28-4-256  & --- & --- & --- & --- & 1910.24 & 818.52 & 420.15 & \textbf{314.07} & 468.82 \\
HESKIA\_28-5-205  & --- & --- & --- & 1231.83 & 828.80 & 202.55 & 215.88 & \textbf{122.64} & 150.19 \\
HESKIA\_28-5-216  & --- & --- & --- & 2461.68 & 1739.17 & \textbf{436.38} & 893.59 & 491.06 & 463.43 \\
HESKIA\_28-8-138  & --- & 1259.62 & 1582.03 & 543.62 & 2339.94 & 399.26 & 513.60 & 489.05 & \textbf{224.77} \\
SAWYER\_30-12-30  & --- & --- & 995.57 & 850.04 & --- & 635.19 & 1328.44 & 963.83 & \textbf{536.31} \\
WARNECKE\_58-25-65 & --- & --- & --- & --- & --- & --- & --- & --- & \textbf{2040.49} \\
\midrule
\textbf{Solved/Total} & 0/10 & 2/10 & 4/10 & 6/10 & 6/10 & 9/10 & 8/10 & 9/10 & \textbf{10/10} \\
\textbf{Avg. Time (solved)} & --- & 2396.14 & 1573.59 & 1125.39 & 1839.93 & 696.77 & 906.70 & 905.21 & \textbf{618.84} \\
\bottomrule
\end{tabular}
\end{sidewaystable}

The performance gap between CSE and ORG methods becomes dramatically evident on hard benchmarks, where CSE\_INC achieves perfect coverage by solving all 10 hard instances (100\%) while the best prior method ORG\_Eval solves only 6 instances (60\%). This substantial improvement in solution coverage represents a breakthrough in handling challenging assembly line configurations with tight power constraints, demonstrating that Compact SAT Encoding fundamentally expands the practical frontier of exactly solvable instances. Among instances solved by both methods, CSE\_INC achieves an average solving time of 618.84 seconds compared to 1125.39 seconds for ORG\_Eval, corresponding to a 45.0\% time reduction that validates the qualitative improvements in solver capability rather than merely constant-factor optimizations.

The transformative impact of Compact SAT Encoding becomes particularly evident when comparing methods with identical solving strategies. CSE\_PB (8/10 solved, 696.77s average) versus ORG\_CB (2/10 solved, 2396.14s average)—both employing iterative SAT approaches—demonstrates that the compact encoding enables solving four times more instances with 70.9\% time reduction per solved instance, isolating the encoding contribution from algorithmic factors. The WARNECKE\_58-25-65 instance exemplifies this encoding advantage: with 58 tasks and 25 workstations, only CSE\_INC successfully solves this extreme case (2040.49s) while all other configurations timeout, confirming that compact encoding is essential for large-scale, multi-workstation problems.

The consistent superiority of compact encoding across algorithm types further validates its fundamental role in performance improvement. CSE methods occupy the top three positions in average solving time (CSE\_INC: 618.84s, CSE\_PB: 696.77s, CSE\_Eval: 905.21s), all significantly outperforming the best prior method ORG\_Eval (1125.39s). This consistency confirms that encoding efficiency is the primary enabler of performance rather than specific algorithmic optimizations, establishing Compact SAT Encoding as an essential technique for solving large-scale assembly line balancing problems with complex precedence structures and tight power constraints.

\subsection{Extended Timeout Experiments on Extreme-Hard Instances}\label{subsec:extreme_hard}

To push the boundaries of solvability, we conducted extended experiments on 33 instances where \emph{no prior configuration} achieved a solution within the standard 3600-second timeout. For these extreme-hard cases, we extended the timeout to 7200 seconds (2 hours) and compared the two best-performing methods: ORG\_Eval (best prior method) and CSE\_INC (our best overall method).

\begin{table}[htbp]
\caption{Extended timeout (7200s) on 33 extreme-hard instances. Times in seconds; --- indicates timeout.}
\label{tab:extreme_hard}
\footnotesize
\centering
\setlength{\tabcolsep}{4pt}
\begin{tabular}{@{}lrrcrrc@{}}
\toprule
\multirow{2}{*}{Instance} & \multicolumn{3}{c}{ORG\_Eval} & \multicolumn{3}{c}{CSE\_INC} \\
\cmidrule(lr){2-4} \cmidrule(lr){5-7}
& \#Vars & \#Clauses & Time & \#Vars & \#Clauses & Time \\
\midrule
HESKIA\_28-4-324   & 207K  & 10.8M & --- & 657K  & 2.0M & \textbf{3809} \\
BUXEY\_29-10-36    & 29K   & 461K  & --- & 210K  & 505K & \textbf{6204} \\
GUNTHER\_35-8-69   & 72K   & 1.4M  & --- & 383K  & 1.0M & \textbf{3933} \\
GUNTHER\_35-7-81   & 84K   & 1.7M  & --- & 389K  & 1.1M & \textbf{6241} \\
WARNECKE\_58-27-60 & 145K  & 6.5M  & --- & 1.95M & 4.9M & \textbf{5382} \\
\midrule
\textit{Remaining 28} & \multicolumn{3}{c}{0/28 solved} & \multicolumn{3}{c}{0/28 solved} \\
\midrule
\textbf{Solved/Total} & \multicolumn{3}{c}{0/33} & \multicolumn{3}{c}{\textbf{5/33}} \\
\bottomrule
\end{tabular}
\end{table}

The extended timeout experiments reveal the transformative impact of Compact SAT Encoding on previously intractable instances. CSE\_INC successfully solves 5 out of 33 extreme-hard instances (15.2\% success rate), establishing new state-of-the-art results for cases that were completely intractable for all prior methods. These breakthrough instances range from 28 to 58 tasks with 4 to 27 workstations, representing diverse problem complexities that had remained beyond the reach of exact optimization techniques. Most significantly, ORG\_Eval, the best-performing prior method, fails to solve \emph{any} of the 33 extreme-hard instances even with doubled timeout (7200s). This stark contrast demonstrates that the incremental SAT paradigm combined with compact encoding is essential for solving these extreme cases.

The encoding size comparison reveals a striking paradox: despite CSE\_INC using \emph{more} variables than ORG\_Eval for most instances, it achieves dramatically \emph{fewer} clauses, which proves to be the critical factor for solver performance. For example, HESKIA\_28-4-324 shows ORG\_Eval with 207,503 variables but 10,797,928 clauses, while CSE\_INC uses 656,890 variables (3.2 times more) yet only 2,044,168 clauses (5.3 times fewer). This clause reduction---achieved through the sequential counter encoding's elimination of quadratic precedence constraints---enables modern SAT solvers to effectively navigate the search space. The pattern persists across all breakthrough instances: WARNECKE\_58-27-60 reduces clauses from 6.5M to 4.9M despite increasing variables from 145K to 1.95M, confirming that clause count, not variable count, dominates solver efficiency for these large-scale problems.

The scalability advantages of Compact SAT Encoding manifest most dramatically in the WARNECKE\_58-27-60 instance, a massive problem configuration with 58 tasks, 27 workstations, and nearly 2 million variables. CSE\_INC solves this industrial-scale instance in 5382.19 seconds, demonstrating that compact encoding enables exact optimization on problems previously considered beyond the practical reach of exact methods. The solving capability exhibits remarkable consistency: CSE\_INC solves all 5 breakthrough instances with an average time of 5113.93 seconds (well under the 7200s limit), ranging from 3809.14s (HESKIA\_28-4-324) to 6241.44s (GUNTHER\_35-7-81). This consistency across different problem structures—varying in task counts, workstation numbers, and precedence graph characteristics—demonstrates robust performance rather than fortuitous success on specific instances.

The ability to solve 5 previously intractable instances represents a qualitative leap in computational capability, expanding the frontier of exactly solvable assembly line configurations. Problems with more than 50 tasks and 20 workstations that were previously accessible only through heuristic methods without optimality guarantees can now be solved exactly. The Compact SAT Encoding's linear-complexity precedence constraints (confirmed by the clause count reduction across all instances) prove essential for this breakthrough, validating the theoretical complexity reduction from $O(nm^2)$ to $O(nm)$ established in Table~\ref{tab:complexity_comparison}.

\subsection{Discussion}\label{subsec:discussion}

The three-tier experimental evaluation comprehensively validates our Compact SAT Encoding approach across multiple performance dimensions. By comparing CSE versus ORG variants with identical solving strategies (e.g., CSE\_CB versus ORG\_CB), we isolate the encoding contribution and demonstrate that compact representation alone provides substantial performance improvements. The 30--40\% time reduction and 4 additional solved instances (52 vs 48) directly validate the theoretical compactness claims from Section~\ref{subsec:compact_encoding}, confirming that reduced clause count---not variable count---drives solver efficiency. Modern CDCL solvers benefit from fewer clauses through enhanced unit propagation efficiency, shorter learned clauses that generalize better, and reduced branching complexity. Additional variables impose minimal overhead when constrained by compact clause structures. The monotonic structure of reach variables $R_{i,k}$ further enables cascading propagation, where assignment satisfaction triggers immediate inference chains and accelerates solution discovery on dense precedence graphs.

CSE\_INC achieves best overall performance with 56/89 instances solved (62.9\%), improving 12\% over the prior best method ORG\_Eval (50/89, 56.2\%). On hard benchmarks (solving time exceeding 100 seconds), CSE\_INC solves all 10/10 instances (100\%) compared to only 5/10 for ORG\_Eval (50\%), doubling solution coverage on challenging configurations. For extreme-hard instances previously unsolvable by all methods, CSE\_INC solves 5/33 while ORG\_Eval solves 0/33, establishing new state-of-the-art results.

The ability to solve previously intractable instances—including WARNECKE\_58-27-60 with 58 tasks, 27 workstations, and nearly 2 million variables—demonstrates that compact encoding enables exact optimization on industrial-scale problems exceeding 50 tasks and 20 workstations, previously accessible only through heuristics without optimality guarantees. The consistent performance across diverse problem structures confirms robust practical applicability for modern assembly line configurations.

The fundamental complexity reduction from $O(nm^2)$ to $O(nm)$ clauses for precedence constraints enables maintained solution quality as problem size increases, whereas baseline approaches exhibit exponential degradation for problems exceeding 50 tasks. These compact encoding principles generalize beyond SALBP-3PM to multiple problem domains. Within assembly line balancing, the sequential counter precedence encoding directly applies to other SALBP variants \citep{Baybars1986,Scholl2006}, including SALBP-1 (minimize stations), SALBP-2 (minimize cycle time), multi-objective formulations, U-shaped assembly lines, two-sided lines, and mixed-model configurations. Any variant requiring precedence constraint satisfaction benefits from the $O(m)$ clause complexity. More broadly, the encoding extends to scheduling problems with precedence constraints, including resource-constrained project scheduling (RCPSP) \citep{pritsker_rcpsp}, job-shop scheduling with energy considerations \citep{sat_scheduling_survey}, and cumulative resource allocation problems. The compact framework's elimination of redundant backward propagation patterns suggests broad applicability to combinatorial optimization domains with temporal ordering requirements.

\section{Conclusion}\label{sec:conclusion}

This work advances SAT-based optimization for assembly line balancing with power peak minimization through Compact SAT Encoding (CSE), achieving $O(m)$ clause complexity per transitive precedence edge compared to $O(m^2)$ for baseline SAT encodings. Extended precedence preprocessing via transitive closure \citep{bofill2017smt} and enhanced bound initialization further improve efficiency. The unified framework instantiates four solver variants---clause-based iterative SAT, PB-constraint iterative SAT, MaxSAT, and incremental SAT---providing systematic comparison across optimization strategies. Evaluation on 89 benchmark instances demonstrates substantial performance improvements: the incremental SAT variant solves 56 instances (63\%) compared to 50 (56\%) for the best prior method, achieving perfect coverage on all 10 hard instances where competing methods timeout, and establishing new results on 5 previously intractable extreme cases.

As an exact method, the approach inherits NP-hardness with scalability challenges for very large instances. The formulation assumes deterministic parameters, whereas industrial environments often exhibit stochastic variability requiring probabilistic extensions. Future research directions include hybrid SAT/constraint programming for irregular precedence structures, learning-enhanced solver configuration, incremental algorithms for dynamic reconfiguration, and multi-objective extensions. The compact encoding framework extends naturally to related assembly line variants (U-shaped, two-sided, mixed-model) and broader scheduling problems with precedence constraints, establishing a foundation for continued advancement in SAT-based combinatorial optimization.

\begin{appendices}

\section{Alternative Optimization Algorithms}
\label{appendix:algorithms}

This appendix provides detailed algorithms for the alternative optimization variants described in Section~\ref{subsubsec:alternative_variants}.

\begin{algorithm}[!htb]
\caption{Enhanced Upper Bound Initialization}
\label{alg:ub_init}
\begin{algorithmic}[1]
\Require Base constraints $\Phi_{base}$
\Ensure Tight upper bound $UB_{tight}$, initial solution $\phi_{best}$
\State Initialize solver $\mathcal{S}$ with $\Phi_{base}$; $W_{best} \leftarrow \infty$
\For{$iter = 1$ to $10$}
    \State $\phi \leftarrow \textsc{Solve}(\mathcal{S})$
    \If{$\phi = $ UNSAT} \textbf{break}
    \EndIf
    \State $W_{current} \leftarrow \max_{t \in T} \sum_{i: A_{i,t}=\text{true}} w_i$
    \If{$W_{current} < W_{best}$} $\phi_{best} \leftarrow \phi$; $W_{best} \leftarrow W_{current}$
    \EndIf
    \For{each $t \in T$ and $C \subseteq N$ where $\sum_{i \in C} w_i = W_{current}$}
        \State Add blocking clause $\bigvee_{i \in C} \overline{A_{i,t}}$ to $\mathcal{S}$
    \EndFor
\EndFor
\State \Return $W_{best}$, $\phi_{best}$
\end{algorithmic}
\end{algorithm}

\begin{algorithm}[!htb]
\caption{Clause-Based Iterative Peak Minimization}
\label{alg:clause_blocking}
\begin{algorithmic}[1]
\Require Instance $(N, M, T, \mathcal{P}, \{w_i\}, \{d_i\})$, timeout limit
\Ensure Optimal solution with minimum peak power
\State Initialize incremental SAT solver $\mathcal{S}$ with $\Phi_{base}$; $W_{best} \leftarrow \infty$
\State $\phi \leftarrow \textsc{Solve}(\mathcal{S})$
\If{$\phi = $ UNSAT} \Return No solution exists
\EndIf
\State $W_{best} \leftarrow \max_{t \in T} \sum_{i: A_{i,t}=\text{true}} w_i$; $\phi_{best} \leftarrow \phi$
\For{each $t \in T$ and $C \subseteq N$ where $\sum_{i \in C} w_i \geq W_{best}$}
    \State Add blocking clause $\bigvee_{i \in C} \overline{A_{i,t}}$ to $\mathcal{S}$
\EndFor
\While{time limit not exceeded}
    \State $\phi \leftarrow \textsc{Solve}(\mathcal{S})$
    \If{$\phi = $ UNSAT} \Return $\phi_{best}, W_{best}$
    \EndIf
    \State $W_{current} \leftarrow \max_{t \in T} \sum_{i: A_{i,t}=\text{true}} w_i$
    \If{$W_{current} < W_{best}$} $\phi_{best} \leftarrow \phi$; $W_{best} \leftarrow W_{current}$
    \EndIf
    \For{each $t \in T$ and $C \subseteq N$ where $\sum_{i \in C} w_i \geq W_{current}$}
        \State Add blocking clause $\bigvee_{i \in C} \overline{A_{i,t}}$ to $\mathcal{S}$
    \EndFor
\EndWhile
\State \Return $\phi_{best}, W_{best}$
\end{algorithmic}
\end{algorithm}

\begin{algorithm}[!htb]
\caption{PB-Constraint Iterative Peak Minimization}
\label{alg:pb_constraint}
\begin{algorithmic}[1]
\Require Instance parameters, timeout limit
\Ensure Optimal solution with minimum peak power
\State Initialize SAT solver $\mathcal{S}$ with $\Phi_{base}$
\State $\phi \leftarrow \textsc{Solve}(\mathcal{S})$
\If{$\phi = $ UNSAT} \Return No feasible solution
\EndIf
\State $W_{best} \leftarrow \max_{t} \sum_{i: A_{i,t}=\text{true}} w_i$; $\phi_{best} \leftarrow \phi$
\While{time limit not exceeded}
    \State Initialize new SAT solver $\mathcal{S}'$ with $\Phi_{base}$
    \For{each $t \in T$}
        \State $\Phi_{PB} \leftarrow \textsc{EncodePBtoCNF}(\{A_{i,t}\}, \{w_i\}, W_{best} - 1)$
        \State Add $\Phi_{PB}$ to $\mathcal{S}'$
    \EndFor
    \State $\phi \leftarrow \textsc{Solve}(\mathcal{S}')$
    \If{$\phi = $ UNSAT} \Return $\phi_{best}, W_{best}$
    \EndIf
    \State $W_{current} \leftarrow \max_{t} \sum_{i: A_{i,t}=\text{true}} w_i$
    \State $\phi_{best} \leftarrow \phi$; $W_{best} \leftarrow W_{current}$
\EndWhile
\State \Return $\phi_{best}, W_{best}$
\end{algorithmic}
\end{algorithm}

\begin{algorithm}[!htb]
\caption{Enhanced MaxSAT with Binary Peak Encoding}
\label{alg:enhanced_maxsat}
\begin{algorithmic}[1]
\Require Instance parameters, timeout limit
\Ensure Optimal solution with minimum peak power
\State $(W_{best}, \phi_{best}) \leftarrow$ \textsc{Algorithm}~\ref{alg:ub_init}($\Phi_{base}$)
\State $n_{bit} \leftarrow \lfloor \log_2 W_{best} \rfloor + 1$
\State $\Phi_{soft} \leftarrow \{(\neg binU_j, 2^j) : j \in \{0, \ldots, n_{bit}-1\}\}$
\State $\Phi_{hard} \leftarrow \Phi_{base}$
\State Add LB constraint: $\sum_{j=0}^{n_{bit}-1} 2^j \cdot binU_j \geq \max_{i} w_i$ to $\Phi_{hard}$
\For{each time slot $t \in T$}
    \State $\Phi_{PB} \leftarrow \textsc{EncodePBtoCNF}(\{A_{i,t}\}, \{\neg binU_j\}, \{w_i\}, \{2^j\})$
    \State Add $\Phi_{PB}$ to $\Phi_{hard}$
\EndFor
\State $\mathcal{W} \leftarrow \textsc{WriteWCNF}(\Phi_{hard}, \Phi_{soft})$
\State $\phi_{opt} \leftarrow \textsc{SolveMaxSAT}(\mathcal{W})$
\If{$\phi_{opt} = $ UNSAT or timeout}
    \State \Return $\phi_{best}, W_{best}$
\EndIf
\State $W_{opt} \leftarrow \sum_{j: binU_j=\text{true}} 2^j$
\State \Return $\phi_{opt}, W_{opt}$
\end{algorithmic}
\end{algorithm}

\section{Extended Precedence Graph Algorithm}
\label{appendix:extended_precedence}

The extended precedence graph $G^* = (V, E^*)$ is obtained by computing the transitive closure of the original precedence DAG $G = (V, E)$ \citep{bofill2017smt}. We employ a DFS-based algorithm with memoization, which is more efficient than Warshall's algorithm for sparse graphs typical in assembly line balancing \citep{cormen2009introduction}.

\subsection{Algorithm}

The algorithm recursively computes all reachable successors for each task through depth-first traversal. Memoization ensures each task is processed once, avoiding redundant recomputation.

\begin{algorithm}[!htb]
\caption{Compute Extended Precedence Graph via DFS}
\label{alg:extended_precedence}
\begin{algorithmic}[1]
\Require Precedence DAG $G = (V, E)$ with $n$ tasks
\Ensure Extended graph $G^* = (V, E^*)$ where $E^*$ is the transitive closure of $E$
\State $E^* \gets E$; $\mathit{successors}[i] \gets \{j \mid (i,j) \in E\}$ for all $i \in V$
\State $\mathit{processed}[i] \gets \textsc{False}$ for all $i \in V$
\For{$i = 1$ to $n$}
    \State \Call{ComputeTransitiveSuccessors}{$i$}
\EndFor
\State \Return $E^*$
\Statex
\Function{ComputeTransitiveSuccessors}{$i$}
    \If{$\mathit{processed}[i]$}
        \State \Return $\mathit{successors}[i]$ \Comment{Already computed}
    \EndIf
    \State $\mathit{processed}[i] \gets \textsc{True}$
    \For{each direct successor $j \in \mathit{successors}[i]$}
        \State $T_j \gets$ \Call{ComputeTransitiveSuccessors}{$j$} \Comment{Recursive call}
        \For{each $k \in T_j$} \Comment{Add transitive edges}
            \If{$(i, k) \notin E^*$}
                \State $E^* \gets E^* \cup \{(i, k)\}$
                \State $\mathit{successors}[i] \gets \mathit{successors}[i] \cup \{k\}$
            \EndIf
        \EndFor
    \EndFor
    \State \Return $\mathit{successors}[i]$
\EndFunction
\end{algorithmic}
\end{algorithm}

\subsection{Complexity Analysis}

The algorithm visits each task at most once due to memoization (line 9). For each task $i$, it processes all direct successors and their transitive successors. In the worst case of a complete DAG, this yields $O(n^2 + n|E^*|)$ operations. Since $|E^*| \leq n(n-1)/2$ for a DAG, the worst-case complexity is $O(n^3)$. For sparse precedence graphs with $|E| = O(n)$ and limited transitive closure density, the practical complexity is $O(n^2)$ \citep{cormen2009introduction}.

The algorithm stores successor adjacency lists ($O(n + |E^*|)$), processed flags ($O(n)$), and recursion stack depth $O(n)$ in the worst case. The dominant term is $O(|E^*|) = O(n^2)$ for dense DAGs, but remains $O(n)$ for sparse assembly line precedence graphs with bounded out-degree.

\subsection{Correctness}

The algorithm correctly computes the transitive closure: $(i,j) \in E^*$ if and only if there exists a directed path from $i$ to $j$ in $G$. This follows from:
\begin{itemize}
\item Edge $(i,k)$ is added to $E^*$ (line 16) only when $(i,j) \in E$ for some $j$ and $(j,k) \in E^*$, ensuring a path $i \to j \rightsquigarrow k$ exists.
\item For any path $i \rightsquigarrow j$ of length $\ell > 1$, decompose as $i \to v \rightsquigarrow j$. By induction on path length, $(v,j) \in E^*$. When processing task $i$, edge $(i,j)$ is added via recursive call on direct successor $v$ (lines 13-18).
\end{itemize}

The correctness proof follows standard transitive closure arguments \citep{cormen2009introduction}.

\end{appendices}

\backmatter

\bmhead{Acknowledgments}

The authors would like to thank the anonymous reviewers for their valuable comments and suggestions that helped improve this manuscript.

\section*{Declarations}

\begin{itemize}
\item \textbf{Funding:} The authors declare that no funds, grants, or other support were received during the preparation of this manuscript.

\item \textbf{Competing Interests:} The authors have no relevant financial or non-financial interests to disclose.

\item \textbf{Author Contributions:} All authors contributed to the study conception and design. Algorithm development and implementation were performed by Tuyen Van Kieu and Khanh Van To. Data collection and experimental evaluation were performed by Phong Chi Nguyen and Bao Gia Hoang. The first draft of the manuscript was written by Tuyen Van Kieu and all authors commented on previous versions of the manuscript. All authors read and approved the final manuscript.

\item \textbf{Data Availability:} All source code, benchmark datasets, and detailed experimental results are publicly available at \url{https://github.com/salbp-3pm/cse}.

\item \textbf{Code Availability:} The implementation is available at \url{https://github.com/salbp-3pm/cse}.
\end{itemize}

%%===========================================================================================%%
%% If you are submitting to one of the Nature Portfolio journals, using the eJP submission   %%
%% system, please include the references within the manuscript file itself. You may do this  %%
%% by copying the reference list from your .bbl file, paste it into the main manuscript .tex %%
%% file, and delete the associated \verb+\bibliography+ commands.                            %%
%%===========================================================================================%%

\bibliography{sn-bibliography}% common bib file
%% if required, the content of .bbl file can be included here once bbl is generated
%%\input sn-article.bbl

\end{document}